\documentclass[fleqn,usenatbib]{mnras}

\usepackage{newtxtext,newtxmath}

\usepackage[T1]{fontenc}

\DeclareRobustCommand{\VAN}[3]{#2}
\let\VANthebibliography\thebibliography
\def\thebibliography{\DeclareRobustCommand{\VAN}[3]{##3}\VANthebibliography}

\usepackage{graphicx}	
\usepackage{amsmath}	
\usepackage{hyperref}

\usepackage{scalerel,tikz}
\usetikzlibrary{svg.path}
\definecolor{orcidlogocol}{HTML}{A6CE39}
\tikzset{orcidlogo/.pic={
 \fill[orcidlogocol] svg{M256,128c0,70.7-57.3,128-128,128C57.3,256,0,198.7,0,128C0,57.3,57.3,0,128,0C198.7,0,256,57.3,256,128z};
 \fill[white] svg{M86.3,186.2H70.9V79.1h15.4v48.4V186.2z}
 svg{M108.9,79.1h41.6c39.6,0,57,28.3,57,53.6c0,27.5-21.5,53.6-56.8,53.6h-41.8V79.1z M124.3,172.4h24.5c34.9,0,42.9-26.5,42.9-39.7c0-21.5-13.7-39.7-43.7-39.7h-23.7V172.4z}
 svg{M88.7,56.8c0,5.5-4.5,10.1-10.1,10.1c-5.6,0-10.1-4.6-10.1-10.1c0-5.6,4.5-10.1,10.1-10.1C84.2,46.7,88.7,51.3,88.7,56.8z};
}}
\newcommand\orcidicon[1]{\href{https://orcid.org/#1}{\mbox{\scalerel*{
\begin{tikzpicture}[yscale=-1,transform shape]
\pic{orcidlogo};
\end{tikzpicture}
}{|}}}}

\newcommand{\aref}[1]{\hyperref[#1]{Appendix~\ref{#1}}}

\title{Reconstructing three-dimensional densities from two-dimensional observations of molecular gas}

\author[Hu et al.]{
Zipeng Hu,$^{\orcidicon{0000-0002-3758-552X}\,1}$\thanks{zphu.charles@gmail.com (ZPH)}
Mark R.~Krumholz,$^{\orcidicon{0000-0003-3893-854X}\,1,2}$\thanks{mark.krumholz@anu.edu.au (MRK)}
Christoph Federrath $^{\orcidicon{0000-0002-0706-2306}\,1,2}$\thanks{christoph.federrath@anu.edu.au (CF)}
Riwaj Pokhrel $^{\orcidicon{0000-0002-0557-7349}\,3}$\thanks{riwajpokhrel@gmail.com (RP)}
\newauthor and Robert A. Gutermuth $^{\orcidicon{0000-0002-6447-899X}\,4}$\thanks{rob.gutermuth@gmail.com (RAG)}
\\
$^{1}$Research School of Astronomy and Astrophysics, Australian National University, Canberra, ACT 2611, Australia\\
$^{2}$ARC Centre of Excellence for Astronomy in Three Dimensions (ASTRO-3D), Canberra, ACT 2611, Australia\\
$^{3}$Ritter Astrophysical Research Center, Department of Physics and Astronomy, University of Toledo, Toledo, OH 43606, USA\\
$^{4}$Department of Astronomy, University of Massachusetts, 710 North Pleasant Street, Amherst, MA 01003, USA\\
}

\date{Accepted XXX. Received YYY; in original form ZZZ}

\pubyear{2020}

\begin{document}
\label{firstpage}
\pagerange{\pageref{firstpage}--\pageref{lastpage}}
\maketitle

\begin{abstract}
Star formation has long been known to be an inefficient process, in the sense that only a small fraction $\epsilon_{\rm ff}$ of the mass of any given gas cloud is converted to stars per cloud free-fall time. However, developing a successful theory of star formation will require measurements of both the mean value of $\epsilon_{\rm ff}$ and its scatter from one molecular cloud to another. Because $\epsilon_{\rm ff}$ is measured relative to the free-fall time, such measurements require accurate determinations of cloud volume densities. Efforts to measure the volume density from two-dimensional projected data, however, have thus far relied on treating molecular clouds as simple uniform spheres, while their real shapes are likely to be filamentary and their density distributions far from uniform. The resulting uncertainty in the true volume density is likely to be one of the major sources of error in observational estimates of $\epsilon_{\rm ff}$. In this paper, we use a suite of simulations of turbulent, magnetized, radiative, self-gravitating star-forming clouds in order to examine whether it is possible to obtain more accurate volume density estimates and thereby reduce this error. We create mock observations from the simulations, and show that current analysis methods relying on the spherical assumption likely yield $\sim 0.26$~dex underestimations and $\sim 0.51$~dex errors in volume density estimates, corresponding to a $\sim 0.13$~dex overestimation and a $\sim 0.25$~dex scatter in $\epsilon_{\rm ff}$, comparable to the scatter in observed cloud samples. We build a predictive model that uses information accessible in two-dimensional measurements -- most significantly the Gini coefficient of the surface density distribution -- to produce estimates of the volume density with $\sim 0.3$~dex less scatter. We test our method on a recent observation of the Ophiuchus cloud, and show that it successfully reduces the $\epsilon_{\rm ff}$ scatter.
\end{abstract}




\section{Introduction}
\label{sec:Introduction}

Because of the wide range of physical processes involved, star formation is one of the least understood phenomena in the universe. However, it is also one of the most important, because star formation plays a key role in the evolution of galaxies and sets the initial conditions for planet formation. One major unsolved problem in this field is why star formation is such an inefficient process. For a star-forming region, the depletion time $t_{\rm dep} = M_{\rm gas}/ \dot M_*$ is the ratio of the gas mass and the star formation rate (SFR). It is a characteristic timescale of star formation. By contrast, the natural timescale for a cloud collapsing under its own gravity is the free-fall time,
\begin{equation}
    t_{\rm ff} = \sqrt{\frac{3\pi}{32G\rho}},
	\label{eq:tff}
\end{equation}
where $G$ is the gravitational constant and $\rho$ is the volume density. The star formation efficiency (SFE) , defined as \citep{2005ApJ...630..250K}
\begin{equation}
    \epsilon_{\rm ff} = \frac{t_{\rm ff}}{t_{\rm dep}} = \sqrt{\frac{3\pi}{32G\rho}}\frac{\dot M_*}{M_{\rm gas}},
    \label{eq:star formation efficiency}
\end{equation}
characterises the efficiency of the star formation process. A value of $\epsilon_{\rm ff} \sim 1$ for a given star-forming region indicates that the region is giving birth to stars with little resistance to self-gravity, i.e., all the gas collapses into stars in a single free-fall time. On the contrary, if $\epsilon_{\rm ff}$ is low, this implies that some other process, for example magnetic or turbulent pressure, is obstructing free-fall collapse and impeding star formation \citep{Federrath_2015,Federrath_2018}. 

\citet{Zuckerman74a} were the first to point out that comparing the Milky Way's star formation rate ($\sim 1$ $\rm M_\odot$ yr$^{-1}$), total mass of molecular clouds ($\sim 10^9$ $\rm M_\odot$), and typical molecular cloud free-fall time ($\sim 10$ Myr) implies that molecular clouds have $\epsilon_{\rm ff}\ll 1$, and \citet{Krumholz_Tan_2007} extended this conclusion to the denser parts of molecular clouds traced by molecules such as HCN. \citet{Krumholz2019} summarise more recent observations on both sub-galactic and whole-galaxy scales, and show that these yield $\epsilon_{\rm ff}$ estimates consistent with a near universal value $\epsilon_{\rm ff} \sim 0.01$ \citep[e.g.][]{Heyer2016, Ochsendorf2017, Onus_2018, Utomo2018}. These results have a study-to-study dispersion of $\approx 0.3$~dex, and a dispersion of about 0.3 -- 0.5 dex within any single study.

The origin of the low observed value of $\epsilon_{\rm ff}$ is one of the major puzzles in star formation theory. To explain it, different groups have built models that can be classified into two main types. One group of theorists explain this phenomenon by focusing on galactic scale physical processes \citep[e.g.][]{Kim_2011, 2011ApJ...731...41O, Faucher-Giguere13a}, while others construct their models by summing up star formation in individual molecular clouds, each of which has a small value of $\epsilon_{\rm ff}$ due to some internal regulation process \citep[e.g.][]{1994ApJ...435L.121E,2011ApJ...731...25K,Federrath_2012}. Both classes of models predict similarly-low $\epsilon_{\rm ff}$ values on average, but they differ substantially in their predictions for the \textit{dispersion} of $\epsilon_{\rm ff}$ on sub-galactic scales -- models where star formation is regulated only on galactic scales generally predict much larger dispersions than those where it is regulated on the cloud scale \citep{Lee_2016, Krumholz_2020}. This provides a strong motivation for measuring the cloud-scale distribution of $\epsilon_{\rm ff}$ values with enough fidelity that we can determine not just its mean value, but also its dispersion. Such measurements also offer an invaluable opportunity to test prescriptions for star formation and feedback in large-scale galaxy and cosmological simulations, since different prescriptions for these processes yield differing distributions of $\epsilon_{\rm ff}$ \citep[e.g.,][]{Semenov16a, Grudic19a, Grisdale_2019, Fujimoto19a}. In order to take advantage of this opportunity, however, we must be able to separate true dispersion from observational errors; the more we can decrease observational errors in measurements of $\epsilon_{\rm ff}$, the more we can constrain theoretical models.

Examining \autoref{eq:star formation efficiency}, we can see that the value of $\epsilon_{\rm ff}$ is related to the SFR, gas mass and volume density. All three parameters carry observational uncertainties, but the volume density is the dominant one. While the other two parameters can be obtained from two-dimensional surface density maps, the volume density is an inherently three-dimensional property, estimates of which are inevitably subject to projection effects. The scale of volume density uncertainties depends on the measurement method. One method is to estimate $\rho$ with density-sensitive multiline spectroscopy \citep{2004ApJS..152...63G,Ginsburg_2013, 2017ApJ...835..217L, Onus_2018}, but this is observationally expensive, and requires significant calibration with uncertain theoretical models. A more direct approach is to derive $\rho$ from the column density $\Sigma$ of observed star-forming gas, relying on assumptions about the line-of-sight depth. For extragalactic observations on scales $\gtrsim 100$ pc, one can estimate the depth from consideration of hydrostatic balance within a galactic disc \citep[e.g.,][]{Utomo2018}, but this approach is not available for surveys focusing on nearby molecular clouds on smaller scales, which are the measurements that are most valuable for testing theoretical models.

Instead, the most common approach in the literature is to assume the cloud being observed is approximately spherical, so its depth along the line sight is comparable to its size in the plane of the sky. \cite{Heyer2016}, for example, identify dense clumps in ATLASGAL dust maps, and for each clump they measure the total area $A$ and total mass $M_{\rm cloud}$. From these two they compute the mean surface density $\bar{\Sigma}=M_{\rm cloud}/A$ and assign a mean radius $R_{\rm eff} = \sqrt{A/\pi}$. Therefore, under the spherical assumption the spherical volume density $\rho_{\rm sph}$ is simply
\begin{equation}
    \rho_{\rm sph} = \frac{3M_{\rm cloud}}{4\pi R_{\rm eff}^3} = \frac{3\bar{\Sigma}}{4\sqrt{A/\pi}}.
    \label{eq:rhosph}
\end{equation}
A number of other authors have used the same basic approach in the Milky Way \citep[e.g.][]{Krumholz2012, 2013ApJ...778..133L, Evans_2014, Pokhrel2021} and in the Large Magellanic Cloud \citep{Ochsendorf2017}.

However, the errors and biases that result from the spherical assumption are at present poorly understood. For decades, filamentary structures have been observed to be a common feature of the interstellar medium (ISM) \citep[e.g.][]{1979ApJS...41...87S, 2005PASJ...57S...1D,Arzoumanian_2011, Andr__2014, 2016A&A...586A..27K}. Contours identified on the surface density map of these structures are elongated. Thus, the volume density derived under the spherical assumption may be quite different from the true mean density. Moreover, even for molecular clouds with perfectly spherical shapes, the mean volume density may still not reflect the mean free-fall time of the whole region. As shown in \autoref{eq:tff}, $t_{\rm ff} \propto \rho^{-0.5}$, which is a non-linear correlation. Thus, if the molecular cloud has a non-uniform mass distribution (which is very likely), the value of $t_{\rm ff}$ determined by integrating sub-regions would not be equal to the value calculated with the mean density of the whole region \citep[e.g.][]{Hennebelle2011,Federrath_2012,Federrath2013,Salim2015}.

Given the importance of volume density measurements and the potential problems of the commonly-used spherical assumption, our goal is to find an improved method to estimate the three dimensional (3D) volume density from two dimensional (2D) observations. Since the true value of volume density can only be determined with 3D data, we turn to numerical simulations, from which we can obtain all 3D properties of the simulated molecular clouds. Using these simulations, we generate mock observations and place surface density contours over them. For each contour, we calculate the true volume density, together with a number of other parameters (mean surface density, velocity dispersion, mass of enclosed stars, etc.) that would be accessible in realistic 2D observations. We use these data to both calibrate the expected error in estimates of $\epsilon_{\rm ff}$ that rely on the spherical assumption, and to develop a predictive model for the volume density that can be used to reduce this error.

This paper is structured as follows. \autoref{sec: Simulations and analysis methods} summarises the simulation data and the data analysis methods. \autoref{sec: Results} presents the results of the analysis and the predictive model. \autoref{sec: Discussion} discusses the physical meaning behind the proposed model and telescope beam effects. \autoref{sec: Analysis on Ophiuchus Cloud} presents a sample application of our model to recent observations of the Ophiuchus cloud, while \autoref{sec: future work} discusses possible future work in this area. \autoref{sec: Conclusion} concludes the work done in the paper.

\section{Simulations and analysis methods}
\label{sec: Simulations and analysis methods}

The simulations we use in our study are from the work of \citet[hereafter C18]{Cunningham2018}. We choose these simulations because they include detailed treatments of many physical processes: gravity, magnetic fields, turbulence, mechanical jets/outflows, and radiation feedback. Moreover, these simulations produce SFEs and initial mass function (IMF) peaks that are both stable in time and are close matches to recent observations, and they span a wide range of turbulent and magnetic field characteristics, allowing us to check for systematic variations with these properties. We start this section with a brief introduction to the main features of the C18 simulations, and then describe the data analysis methods we apply in the remainder of this section.

\subsection{Summary of simulations}
\label{sec: Physics included and numerical method}

C18 uses the ORION2 adaptive mesh refinement (AMR) code \citep{Li_2012}. It solves the equations of ideal magnetohydrodynamics (MHD) using the scheme of \cite{2012ApJS..198....7M}, together with coupled self-gravity \citep{1998ApJ...495..821T, Klein_1999} and radiation transfer \citep{Krumholz_2007}. The C18 simulations include driven turbulence, produced following the driving recipe of \cite{Mac_Low_1999}. They include protostellar outflows following the procedure described in \cite{Cunningham_2011}; star formation follows the sink particle algorithm of \cite{Krumholz_2004}, while protostar evolution and radiative feedback use the model developed by \cite{Offner_2009}. We refer readers to C18 for full details on how each of these physical processes are implemented.

C18 includes nine individual simulations with slightly different initial conditions, whose properties are summarised in \autoref{tab:simulation parameters}. For all simulations, the AMR hierarchy is initialized on a $256^3$ base grid denoted as $\mathcal{L} = 0$. The highest refinement level $\mathcal{L}_{\rm max}$ = 4 (so the highest resolution is 1/$2^4$ times of that of the base grid) for three simulations and $\mathcal{L}_{\rm max}$ = 3 for the other six. The values of $\mathcal{L}_{\rm max}$ for each simulation are listed in \autoref{tab:simulation parameters}. The initial state consists of molecular gas with solar metallicity, a mean molecular weight 2.33 $m_{\rm p}$ and an initial temperature $T_{\rm g}$ = 10~K. Thus, the initial sound speed is $c_{\rm s}$ = 0.19 km $\rm s^{-1}$. The simulation domain is a periodic box with size $L$ = 0.65~pc and mean density $\bar{\rho} = 4.46 \times 10^{-20} \: \rm g \: cm^{-3}$, which corresponds to a total mass of $M = 185 \: \rm M_{\odot}$. Therefore, the length scale and the mean density of these simulations represent isolated globules, dense clumps, or filaments within clouds instead of the whole cloud.

In all cases the simulations begin with a uniform medium, and are run for two box crossing times with gravity disabled and turbulent driving turned on, so that the turbulence reaches a statistical steady state. After that point, gravity is turned on; in half the simulations driving continues, while in the other half it is disabled at this point, so that turbulence decays freely. In addition to this variation in driving, the simulations vary in their degree of magnetisation. All simulations begin with a uniform magnetic field, whose strength is parameterised in terms of the mass-to-flux ratio normalised to the critical value, $\mu_{\Phi} = M/M_\Phi$, where $\Phi$ is the magnetic flux through the simulation domain and $M_\Phi = \Phi/2\pi\sqrt{G}$ is the magnetic critical mass \citep{1976ApJ...210..326M}. The C18 simulations include cases with $\mu_{\Phi}$ = 1.56, 2.17, 23.1 and $\infty$ (i.e., no magnetic fields), each run with driving turned on and off, for a total of eight models. In addition, one of the non-driven models is run with protostellar outflows disabled, yielding a total of nine cases.

\begingroup
\setlength{\tabcolsep}{10pt}
\begin{table*}
    \centering
    \begin{tabular}{ccccccc}
    \hline
    Name & $\mu_{\Phi}$ & Outflows & Driving & $\mathcal{L}_{\rm max}$ & Time (Myr) & $M_*/M$  \\
    \hline
    lo & 1.56 & \checkmark & $\times$ & 3 & 1.911 & 0.10\\
    loDrive & 1.56 & \checkmark & \checkmark & 3 & 1.843 & 0.070    \\
    loNW & 1.56 & $\times$ & $\times$ & 3 & 1.640 & 0.13\\
    lo2 & 2.17 & \checkmark & $\times$ & 4 & 1.547 & 0.057    \\
    lo2Drive & 2.17 & \checkmark & \checkmark & 3 & 1.824 & 0.080 \\
    hi & 23.1 & \checkmark & $\times$ & 4 & 1.390 & 0.060\\
    hiDrive & 23.1 & \checkmark & \checkmark & 3 & 1.535 & 0.034 \\
    hydro & $\infty$ & \checkmark & $\times$ & 4 & 1.319 & 0.052\\
    hydroDrive & $\infty$ & \checkmark & \checkmark & 3 & 1.505 & 0.052\\
    \hline
    \end{tabular}
    \caption{The short names and main differences of all 9 simulations in C18. The 1st column is the name of each simulation. The 2nd column is the mass-to-flux ratio normalised to the critical value ($\mu_{\Phi}$). The 3rd and 4th columns indicate whether protostellar outflows and turbulent driving are included in the simulation. The 5th column shows the highest refinement level $\mathcal{L}_{\rm max}$, which is related to the maximum linear resolution by $\Delta x = (524\mbox{ AU})/2^{\mathcal{L}_{\rm max}}$. The 6th column is the simulation time of the snapshot we use for our analysis, with $t=0$ corresponding to the time at which gravity is turned on, and the 7th column is the ratio between the total mass of sink particles and total mass inside the simulation box.}
    \label{tab:simulation parameters}
\end{table*}
\endgroup

\subsection{Data analysis methods}
\label{sec: Data analysis methods}

For the analysis in this paper, we use only the last snapshots, which are taken at the times listed in Table ~\ref{tab:simulation parameters}; here $t=0$ corresponds to the time at which gravity is switched on. The analysis procedure consists of three steps: creating and selecting contours, measurement of the true 3D volume density, and measurement of 2D contour properties. The details of each step will be illustrated below.

\subsubsection{Creating and selecting contours}
\label{sec: Creating and selecting contours}

The first step of the simulation data analysis is to generate and select surface density contours. We start by making projection maps for every snapshot at the native resolution of the simulations along each of the three cardinal axes, yielding 27 gas column density maps. On each map, we define 30 levels of surface density $\Sigma$, uniformly spaced in logarithm between the mean value of the map, $\bar{\Sigma}$\footnote{Note that $\bar{\Sigma}$ is identical for each projection of a single simulation, but differs between the simulations, because at the snapshots we use, different simulations have converted different fractions of their gas to stars.} and the maximum value $\Sigma_{\rm max}$. We start from $\bar{\Sigma}$ rather than from a lower surface density contour because we want to focus on the high-density regions where star formation occurs. From the smallest to the largest of the determined column density levels, we draw contours on the $\Sigma$ map for each level. The set of closed contours generated by this procedure forms the basic data set we will analyse in the remainder of this work.

To select contours suitable for further analysis, we discard those that fail to meet four conditions. First, we project each contour onto the two axes of the $\Sigma$ map, and measure the lengths $L_1$ and $L_2$ of the one-dimensional (1D) projections on both axes. We only retain contours with $L_1, L_2 < L/2$. The reason is that the C18 simulations use periodic boundary conditions, which makes it hard to define the shape and the centre of mass of the contours that cover a significant fraction of the computational box. Second, we discard contours with a mean radius $R_{\rm eff} = \sqrt{A/\pi} < L/100$, where $A$ is the area enclosed by the contour. As shown in \cite{Federrath_2011}, one needs about 30 pixels across a structure to adequately resolve its internal turbulent motions. Since our simulation maps are either $2048^2$ or $4096^2$ in size, this condition ensures that contours are resolved by a minimum radius of $\approx 20$--$40$ pixels, depending on the maximum resolution of the simulation. Thus this criterion guarantees that the internal structures of the selected contours are well resolved. Third, we retain only contours enclosing at least one sink particle. This selection rule is intended to mimic observations, which usually focus on regions selected around observed protostars. The fourth criteria is to select the most massive contour from the retained ones on each level. The reason is that a large contour on a low $\Sigma$ level may break into several smaller ones on a higher level, making the whole sample biased towards the high $\Sigma$ range. Selecting only one contour each level can avoid this bias, and the most massive contour is more representative than others. With the four criteria above, our 27 maps yield 365 contours suitable for further analysis.

\begin{figure}
    \includegraphics[width = \linewidth]{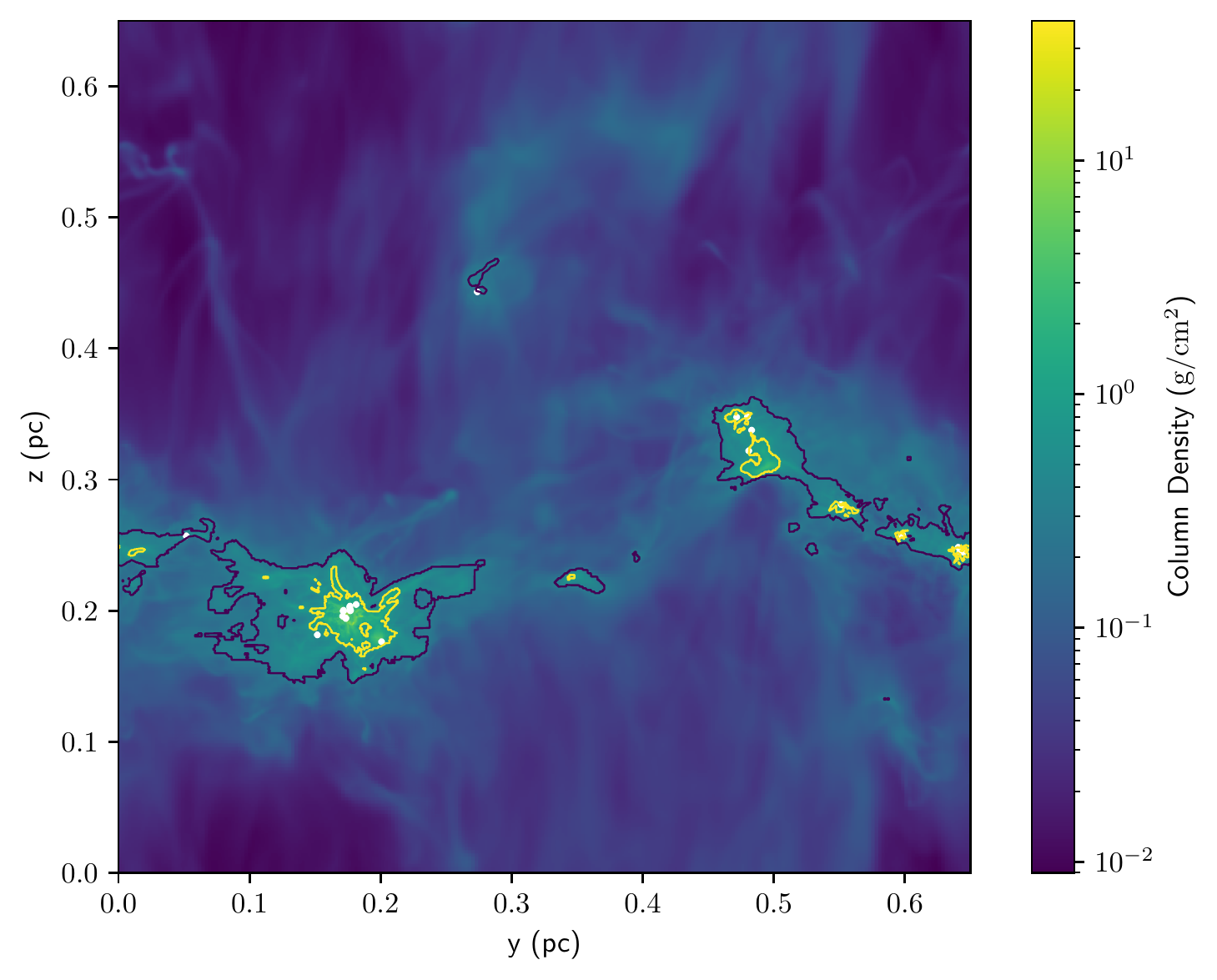}
\caption{The column density map of simulation lo, projected along the $\hat{x}$-axis. The white circles are the projected positions of sink particles. The contours shown represent level~6 (blue; $\Sigma = 0.29 \: \rm g/cm^2 $) and level~11 (yellow; $\Sigma = 0.80 \: \rm g/cm^2 $) of the 30~column density levels determined from the map.} 
\label{fig:lo map}
\end{figure}

We show an example surface density map and contours, in this case for simulation lo projected along the $\hat{x}$-axis, in \autoref{fig:lo map}. The white circles are the projected positions of sink particles, the blue contours are from level 6/30 ($\Sigma = 0.29 \: \rm g/cm^2 $) and the yellow contours are from level 11/30 ($\Sigma = 0.80 \: \rm g/cm^2 $). For reference, the mean surface density of this map is $\bar{\Sigma} = 0.084 \: \rm g/cm^2 $.

\subsubsection{Measuring the effective volume density}
\label{sec: Measurement of rho_eff}

As mentioned in the introduction, for a molecular cloud with a non-uniform mass distribution, a simple mean volume density $\bar{\rho}$ does not reflect the mean free-fall time of the whole cloud, and using it may lead to significant uncertainties when inferring the value of $\epsilon_{\rm ff}$. We therefore define the effective volume density $\rho_{\rm eff}$ to be a free-fall-time-weighted mean density that is more suitable for calculating $\epsilon_{\rm ff}$. For a molecular cloud with non-uniform density and a fixed value of $\epsilon_{\rm ff}$, the SFR is given by \citep{Hennebelle2011,Federrath_2012},
\begin{equation}
    \dot{M}_* = \int \epsilon_{\rm ff} \frac{\rho}{t_{\rm ff}(\rho)}dV = \epsilon_{\rm ff}\sqrt{\frac{32G}{3\pi}} \int \rho^{3/2} dV,
	\label{eq:starFormationRate}
\end{equation}
where the integral is over the cloud volume.
We therefore define the effective free-fall time for the whole mass to be
\begin{equation}
    t_{\rm ff, eff} = \sqrt{\frac{3\pi}{32G\rho_{\rm eff}}},
	\label{eq:tff, eff}
\end{equation}
where $\rho_{\rm eff}$ is our effective volume density, defined implicitly by demanding
\begin{equation}
    \dot{M}_* = \epsilon_{\rm ff} \frac{M_{\rm gas}}{t_{\rm ff,eff}} = \epsilon_{\rm ff} \sqrt{\frac{32G}{3\pi}} \rho_{\rm eff}^{1/2} \int \rho \, dV.
	\label{eq:starFormationRate2}
\end{equation}
Equating \autoref{eq:starFormationRate} and \autoref{eq:starFormationRate2}, we therefore define $\rho_{\rm eff}$ as
\begin{equation}
    \rho_{\rm eff} = \left(\frac{\int \rho^{3/2} \, dV}{\int \rho \, dV}\right)^2,
	\label{eq:effective volume density}
\end{equation}
which is more suitable for calculating $\epsilon_{\rm ff}$ as in \autoref{eq:star formation efficiency}. For each selected contour, we measure $\rho_{\rm eff}$ by evaluating the integrals in \autoref{eq:effective volume density} over a volume defined by the projection of the contour along the line of sight through the full volume of the simulation.

\subsubsection{Measurement of 2D contour properties}
\label{sec: Measurement of 2D contour properties}

\begin{figure}
  \includegraphics[width = \linewidth]{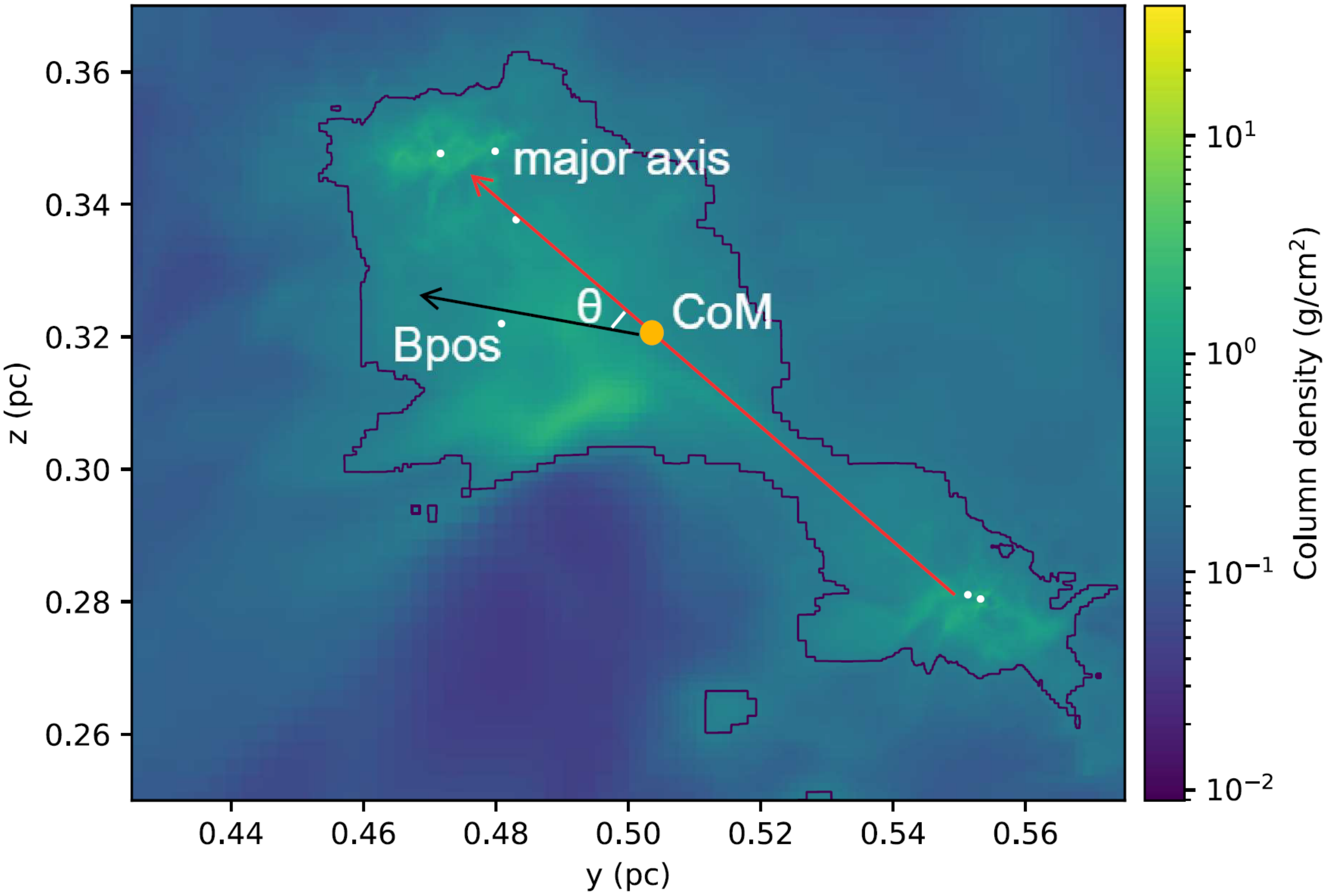}
\caption{Zoom-in on the area around a single contour from \autoref{fig:lo map}; as in that figure, colour shows column density and white points indicate the position of star particles. The orange circle is the position of the centre of mass (CoM) of the contour. The red arrow shows the direction of the major axis, and the black arrow shows the direction of the plane-of-sky magnetic field $B_{\rm pos}$; $\theta$ labels the angle between the major axis and $B_{\rm pos}$.}  
\label{fig:single contour}
\end{figure}

To build a model that can predict $\rho_{\rm eff}$ from 2D observations, we need to determine contour properties that may be related to $\rho_{\rm eff}$. To illustrate our procedure we will use the contour located on the mid-right side of \autoref{fig:lo map} as an example. We zoom in on this contour in \autoref{fig:single contour}, where we show the contour, its centre of mass (CoM) position, major axis direction, and plane-of-sky magnetic field ($B_{\rm pos}$) direction. 

From each selected contour we determine 10 parameters. The 1st parameter is the spherical density $\rho_{\rm sph}$ as defined in \autoref{eq:rhosph}, which we will compare with the value of $\rho_{\rm eff}$ defined in \autoref{eq:effective volume density}. The 2nd is the mean radius of the contour, $R_{\rm eff}$. The 3rd is the ratio between the mean column density of the contour and that of the whole column density map $\Sigma_{\rm contour}/\bar{\Sigma}$; we choose the ratio instead of the absolute value in order to minimise the effect of the difference of gaseous mass between simulations. As stellar feedback may change the mass distribution of a molecular cloud, we select as the 4th parameter the ratio between the total mass of the sink particles inside the contour and the total gas mass of the contour, $M_*/M_{\rm contour}$.

The 5th parameter is the line-of-sight (los) velocity dispersion $\sigma_{v, \rm los}$. We define it as follows, roughly mimicking the way it might be measured from a position-position-velocity data cube using an optically thin tracer: for each pixel $i$ in the projected map that lies inside the contour of interest, we first compute the first moment of the los velocity, $v_{i,\rm los} = \sum_j M_{ij} v_{ij,\rm los} / \sum_j M_{ij}$, where $M_{ij}$ and $v_{ij,\rm los}$ are the mass and los velocity of each cell $j$ along a particular line of sight $i$ through the projected map. We further define the mean los velocity $\bar{v}_{\rm los}$ as the mean of the $v_{i,\rm los}$ values, and the los velocity dispersion by $\sigma_{v,\rm los}^2 = \sum_{ij} (v_{ij,\rm los}-\bar{v}_{\rm los})^2 / N_{\rm p}$, where $N_{\rm p}$ is the total number of pixels included in the contour. Thus, the los velocity dispersion is the root mean square velocity of all computational cells within the contour, measured in the frame where the CoM velocity is zero.

To describe the shapes of our contours, we introduce the ellipticity $e$ as the 6th parameter. The definition of $e$ is
\begin{equation}
    e = 1 - b/a,
	\label{eq:elipticity}
\end{equation}
where $a$ is the semi-major axis length of the contour and $b$ is the semi-minor axis length; $e \sim 0$ corresponds to an extremely elongated contour and $e \sim 1$ describes a nearly circular contour. To determine $a$ and $b$, we first calculate the CoM of the contour. Then for each pixel in the contour with a mass $M_p$ and a displacement from the CoM, $\Delta \mathbf{x} = (\Delta x_{p,1}, \Delta x_{p,2})$, we define the inertia tensor $\mathcal{I}$ as
\begin{equation}
    \mathcal{I}_{ij} = (-1)^{i+j}\sum\limits_p m_p\Delta x_{p,i} \Delta x_{p,j},
	\label{eq:intertia tensor}
\end{equation}
where the sum runs over all pixels interior to the contour. The eigenvalues of $\mathcal{I}$ are $a$ and $b$ (where $a \geq b$ by convention), and the corresponding eigenvectors define the directions of the major and minor axes.

The 7th and 8th parameters are the projected, mass-weighted mean magnetic field strengths in the plane of sky $B_{\rm pos}$ and in the line of sight $B_{\rm los}$; the former is approximately measurable using Zeeman splitting, and the latter using dust polarisation. We define the 9th parameter $\theta$ as the angle between the major axis and $B_{\rm pos}$. For consistency we always choose the smaller angle between the two directions, thus $\theta \in [0, \: \pi/2]$ radian.

\begin{figure}
  \includegraphics[width = \linewidth]{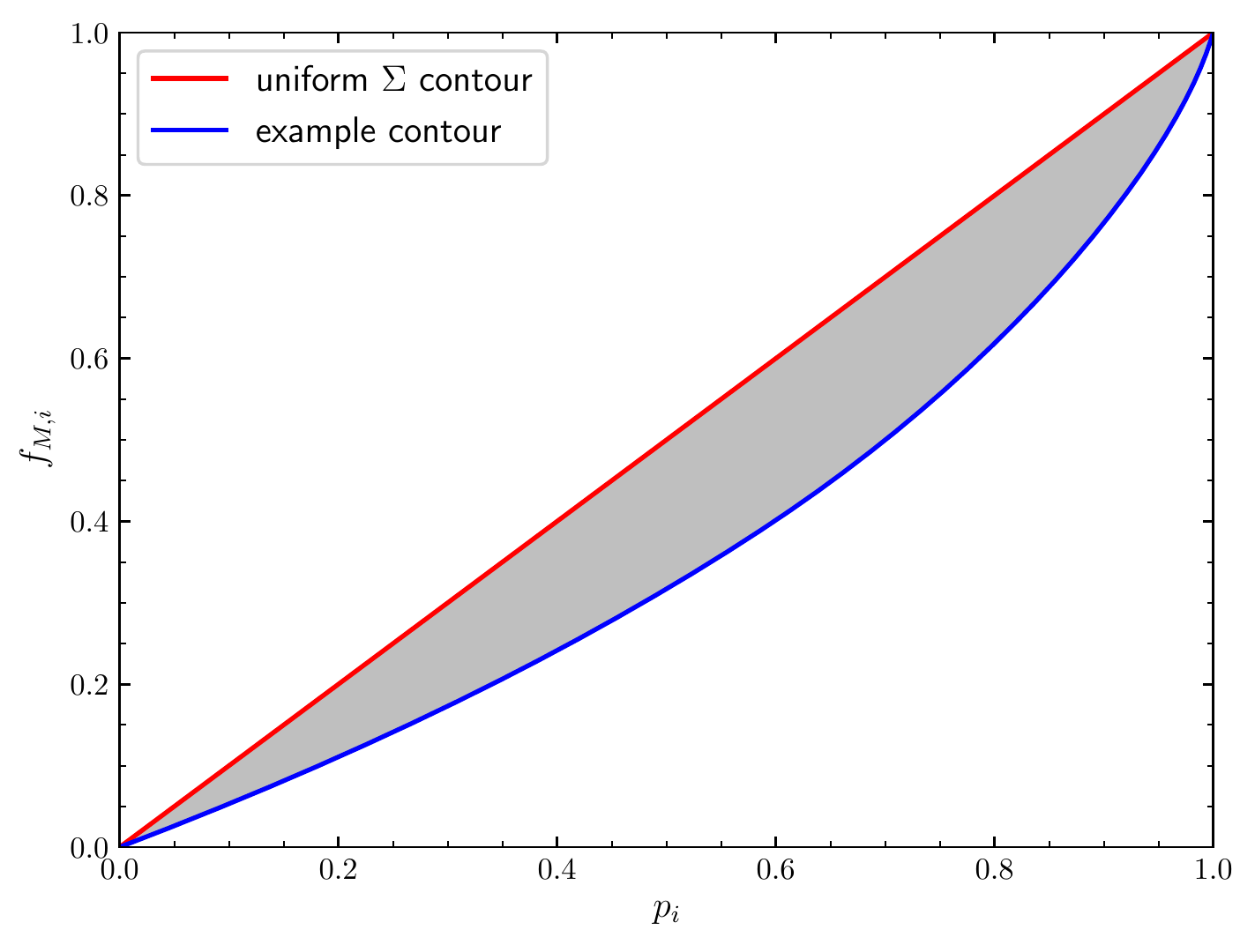}
\caption{Mass fraction $f_{M,i}$ contained in pixels with $\Sigma < \Sigma_i$ as a function of percentile rank $p_i$. The blue line shows this relationship for the example contour shown in \autoref{fig:single contour}, while the red line shows the relationship for a contour with a uniform column density. The Gini coefficient is the ratio of the grey shaded area between the two curves to the area of the right triangle below the uniform density line.
}  
\label{fig:fraction of mass}
\end{figure}

The 10th and last parameter is the Gini coefficient $g$ \citep{Gini_1936} of the column densities of the pixels $\Sigma_i$ enclosed by the contour. To compute this, we first sort the values of enclosed $\Sigma_i$ from the smallest to the largest. For each pixel value $\Sigma_i$, we calculate the fraction of mass $f_{M,i}$ contained in pixels with column density $\Sigma < \Sigma_i$ and plot it against the percentile rank $p_i$ of $\Sigma_i$, i.e., $p_i$ is the fraction of pixels for which $\Sigma < \Sigma_i$. For a contour with constant $\Sigma_i$ (i.e., a uniform column density distribution), $f_M$ is a straight line from (0, 0) to (1, 1). For our example contour whose column density is non-uniform, the start and end of the curve of $f_{M,i}$ versus $p_i$ are the same, but $f_{M,i}$ falls below the one-to-one line for $0<p_i<1$. We show the measured $f_M$ for our example contour, and a hypothetical curve for a uniform column density region, in \autoref{fig:fraction of mass}. The Gini coefficient is defined as the ratio between the area of the gray region and the area of the right triangle under the red curve: formally,
\begin{equation}
    g = 2 \sum_{i=1}^{N_{\rm pix}} \left(p_i - f_{M,i}\right) \left(p_{i+1}-p_i\right),
\end{equation}
where there are $N_{\rm pix}$ pixels within the contour, and by convention $p_{N_{\rm pix}+1} = 1$. Clearly $g$ is bounded to lie between 0 and 1; $g \approx 0$ describes a contour with near uniform surface density, while $g \approx 1$ corresponds to a contour with highly concentrated mass distribution.

\section{Results}
\label{sec: Results}

Having created our sample with the selection of 365 contours and measured the interesting contour properties, we now investigate whether it is possible to build a model that can predict $\rho_{\rm eff}$ from the contour properties. We start by examining the difference between $\rho_{\rm eff}$ and $\rho_{\rm sph}$ in Section~\ref{sec: Comparing rho_eff and rho_sph}. Then we utilize the method of multiple linear fitting to build our model. In the remainder of this section, we describe the effectiveness of our model under different conditions.

\subsection{Comparing $\rho_{\rm eff}$ and $\rho_{\rm sph}$}
\label{sec: Comparing rho_eff and rho_sph}

\begin{figure}
   \includegraphics[width = \linewidth]{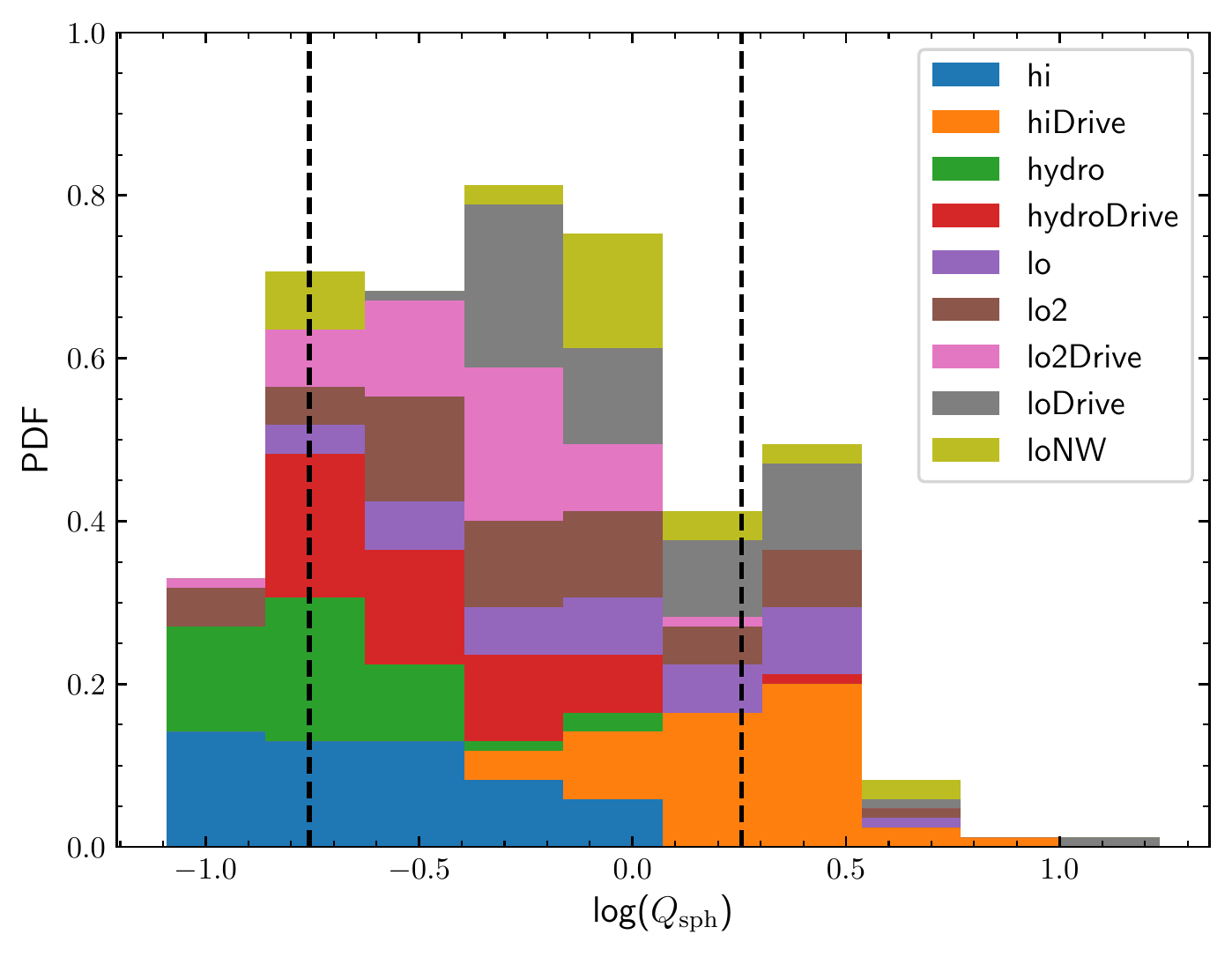}
\caption{Normalized histogram of $\log Q_{\rm sph}$, the quantity that characterises the ratio of the density estimated using the spherical assumption to the true effective density; for example, $\log Q_{\rm sph}$ values of $-1$ and 1 correspond to the spherically-estimated density being too small and too large by a factors of 10, respectively. The simulations from which each measurement of $Q_{\rm sph}$ comes are indicated by colour, as shown in the legend. The two vertical dashed lines show the 16th and 84th percentiles of the distribution.}  
\label{fig:histogram log(rhoSph/rhoEff)}
\end{figure}

For each selected contour, we define
\begin{equation}
Q_{\rm sph} = \frac{\rho_{\rm sph}}{\rho_{\rm eff}}
\end{equation}
as the ratio of the spherical approximation density to the effective density; values of $Q_{\rm sph} > 1$ indicate that the spherical density overestimates the effective density, while values $<1$ indicate underestimates. This will be our figure of merit for the remainder of the paper, i.e., this quantity characterises how well we can approximate the true, 3D density given the projected information to which we have access. A perfect model would yield a distribution of $Q$ values that is a $\delta$ function at $Q=1$. For the 365 selected contours, the mean value of $Q_{\rm sph}$ is $\overline{Q}_{\rm sph} =  0.948$, and the median value is $Q_{\rm sph, med} = 0.544$. We show the full histogram of $\log Q_{\rm sph}$ in \autoref{fig:histogram log(rhoSph/rhoEff)} with the contours' simulation sources labeled. From \autoref{fig:histogram log(rhoSph/rhoEff)} we can see that the distribution of $\log Q_{\rm sph}$ is more weighted to $\log Q_{\rm sph} < 0$, with $\log Q_{\rm sph, med} = -0.26$. The distribution of $\log Q_{\rm sph}$ values varies between individual simulations. Most $\log Q_{\rm sph}$ values for the hydro simulation, for example, are less than 0. To quantify the dispersion of $\log Q_{\rm sph}$, we determine the 16th and 84th percentiles of $\log Q_{\rm sph}$, which we show as black vertical dashed lines in \autoref{fig:histogram log(rhoSph/rhoEff)}. We define the dispersion 
\begin{equation}
\sigma_{\rm sph} \equiv \frac{1}{2} \left(Q_{\rm sph,84} - Q_{\rm sph,16}\right),
\label{eq:sigma_sph_definition}
\end{equation}
where $Q_{\rm sph,16}$ and $Q_{\rm sph,84}$ are the 16th and 84th percentile values, respectively. Thus, for a Gaussian distribution of $\log Q_{\rm sph}$ values, $\sigma_{\rm sph}$ is just the usual Gaussian dispersion. For the data shown in \autoref{fig:histogram log(rhoSph/rhoEff)}, $\sigma_{\rm sph} = 0.51$~dex. Therefore, the volume density determined under the spherical cloud assumption underestimates $\rho_{\rm eff}$ by $\approx  0.26$~dex and carries an uncertainty of $\Delta\rho_{\rm sph} \approx  0.51$~dex.

\subsection{Building the predictive model}
\label{sec: Building the prediction model}

To reduce the uncertainty carried by $\rho_{\rm sph}$, we next build a model to predict the value of $\rho_{\rm eff}$ from 2D contour properties by multiple linear fitting (MLF). As some parameters introduced in Section~\ref{sec: Measurement of 2D contour properties} have wide ranges, we carry out our fits using log-scaled variables. The dependent variable is $Y =  \log(\rho_{\rm eff}/\rho_{\rm sph}) = -\log Q_{\rm sph}$, while the six independent variables are
\begin{equation}
    \mathbf{X} =
    \left[\log R_{\rm eff}, \log\frac{\Sigma_{\rm contour}}{\bar{\Sigma}}, \log \frac{M_*}{M_{\rm contour}}, \log \sigma_{\rm v, los}, e, g\right].
\end{equation}
We omit the magnetic variables for now, because they are not available for the simulations that do not include magnetic fields; we revisit these variables in \autoref{sec: Prediction model with B-field information}. After fitting we obtain the coefficient vector $\mathbf{k}$ and the intercept $b$. Thus, the predicted effective volume density $\rho_{\rm p}$ is
\begin{equation}
    \rho_{\rm p} = C\rho_{\rm sph} = 10^{\mathbf{k}\cdot\mathbf{X} + \,b}\rho_{\rm sph},
	\label{eq:predicted effective volume density}
\end{equation}
where $C \equiv 10^{\mathbf{k}\cdot\mathbf{X}+\,b}$ is the correction factor. By analogy with $Q_{\rm sph}$ and $\sigma_{\rm sph}$ as defined in \autoref{sec: Comparing rho_eff and rho_sph}, we now define $Q_{\rm p} = \rho_{\rm p}/\rho_{\rm eff}$ and $\sigma_{\rm p}$ as the ratio of the predicted and effective densities and half of the distance between the 16th and 84th percentiles of $\log Q_{\rm p}$, respectively. 

\begin{figure}
     \includegraphics[width = \linewidth]{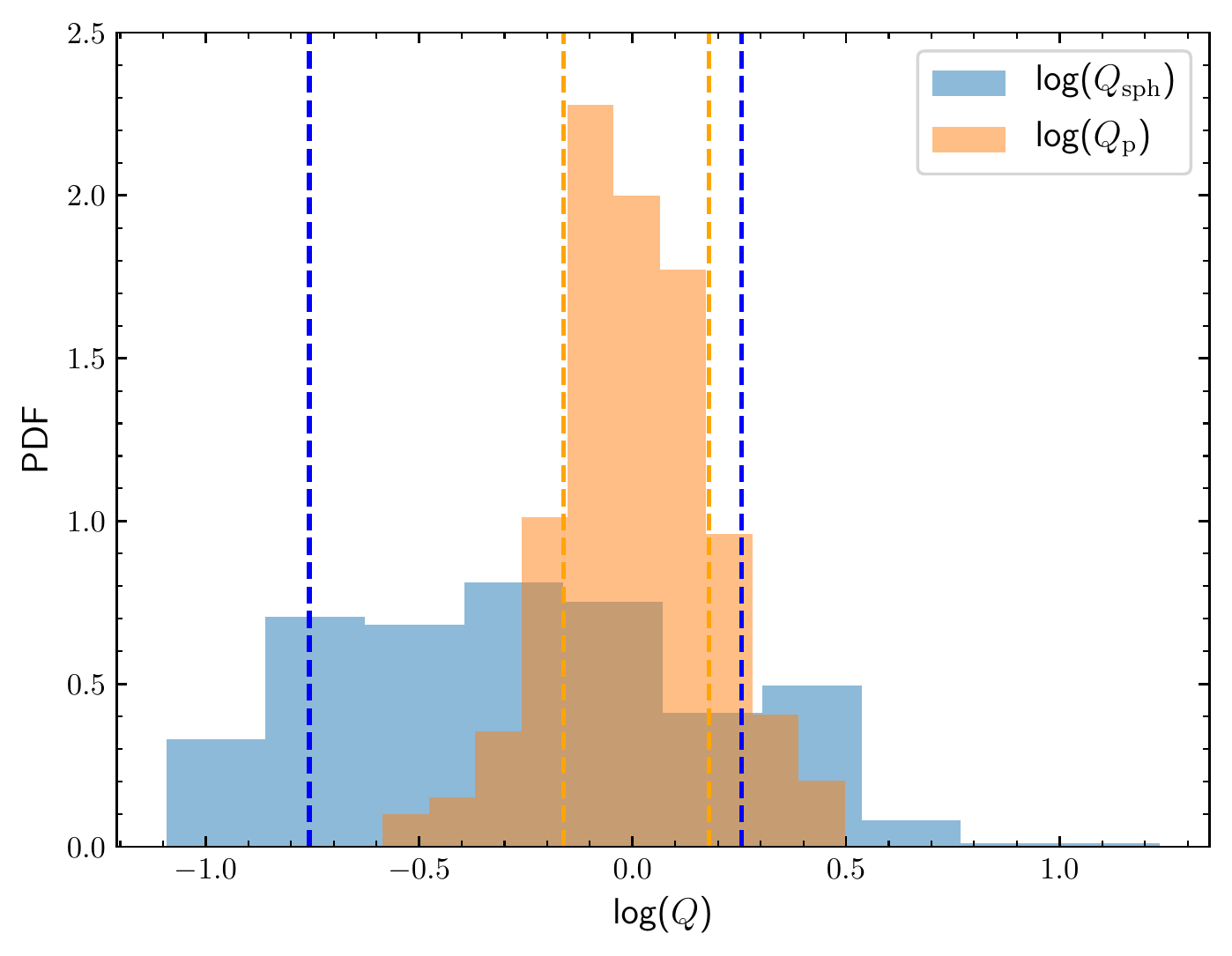}
\caption{Normalized histograms of $\log Q_{\rm sph}$ (blue) and $\log Q_{\rm p}$ (orange). The histogram of $\log Q_{\rm sph}$ values is the same as that shown in \autoref{fig:histogram log(rhoSph/rhoEff)}. The two blue dashed lines show the 16th and 84 percentiles of $\log Q_{\rm sph}$, and the two orange dashed lines show the 16th and 84 percentiles of $\log Q_{\rm p}$. The predictive model substantially reduces the bias and error in estimates of the effective density.}  
\label{fig:histogram overlapped}
\end{figure}

We report the best-fitting values of $\mathbf{k}$ and $b$ as Fit~1 in \autoref{tab:MLF results}. The coefficient of determination for this fit is $R^2 = 0.83$, indicating a strong correlation and justifying our choice of MLF. With the fitted relation, we predict the effective volume density $\rho_{\rm p}$ for every contour in the sample. We compare the normalized histograms of $\log Q_{\rm sph}$ and $\log Q_{\rm p}$ in \autoref{fig:histogram overlapped}. It is obvious that $\log Q_{\rm p}$ is much more narrowly distributed around zero than $\log Q_{\rm sph}$, with $\log Q_{\rm p, med} = 5.4 \times 10^{-4}$. The resulting dispersion, $\sigma_{\rm p} = 0.17$~dex, is also substantially smaller. Thus, this fitted relation not only eliminates the bias, but also reduces the uncertainty in the effective volume density by
\begin{equation}
    \Delta\sigma = \sigma_{\rm sph} - \sigma_{\rm p} = 0.34\mbox{ dex}.
    \label{eq:delta_sigma}
\end{equation}
We use $\Delta\sigma$, the amount by which a given model reduces the scatter in $\log(\rho_{\rm p}/\rho_{\rm eff})$ compared to $\log(\rho_{\rm sph}/\rho_{\rm eff})$, as our figure of merit for evaluating our predictive model from this point forward.

\begingroup
\setlength{\tabcolsep}{10pt} 
\begin{table}
    \centering
    \begin{tabular}{crr}
    \hline
    Quantity & Fit 1 & Fit 2\\
    \hline
    $\log(R_{\rm eff}/\text{pc})$ & 0.47 & 0.52\\
    $\log(\Sigma_{\rm contour}/\bar{\Sigma})$ & 0.16 & $-0.042$\\
    $\log(M_*/M_{\rm contour})$ & 0.042 & 0.031\\
    $\log(\sigma_{\rm v, los}/\text{(cm/s)})$ & $- 0.18$ & $-0.14$\\
    $\log(B_{\rm v, pos}/\text{G})$ & --- & 0.070\\
    $\log(B_{\rm v, los}/\text{G})$ & --- & 0.16\\
    $\theta$ & --- & $-0.0040$\\
    $e$ & 0.055 & 0.21\\
    $g$ & 3.91 & 3.4\\
    \hline
    $b$ & $0.63$ & $1.7$\\
    $R^2$ & 0.83 & 0.87\\
    $\Delta\sigma$ (dex) & 0.34 & 0.33\\
    \hline
    \end{tabular}
    \caption{Results of MLF for the correction factor $C$ between $\rho_{\rm eff}$ and $\rho_{\rm sph}$ (see \autoref{eq:predicted effective volume density}). The top block of rows show the fit coefficients $\mathbf{k}$, and the last three rows provide the intercept $b$, the coefficients of determination $R^2$, and the amount $\Delta\sigma$ by which the fit reduces the dispersion of $\log Q$.}
    \label{tab:MLF results}
\end{table}
\endgroup

\subsection{The effect of including magnetic field data}
\label{sec: Prediction model with B-field information}
\begin{figure}
  \includegraphics[width = \linewidth]{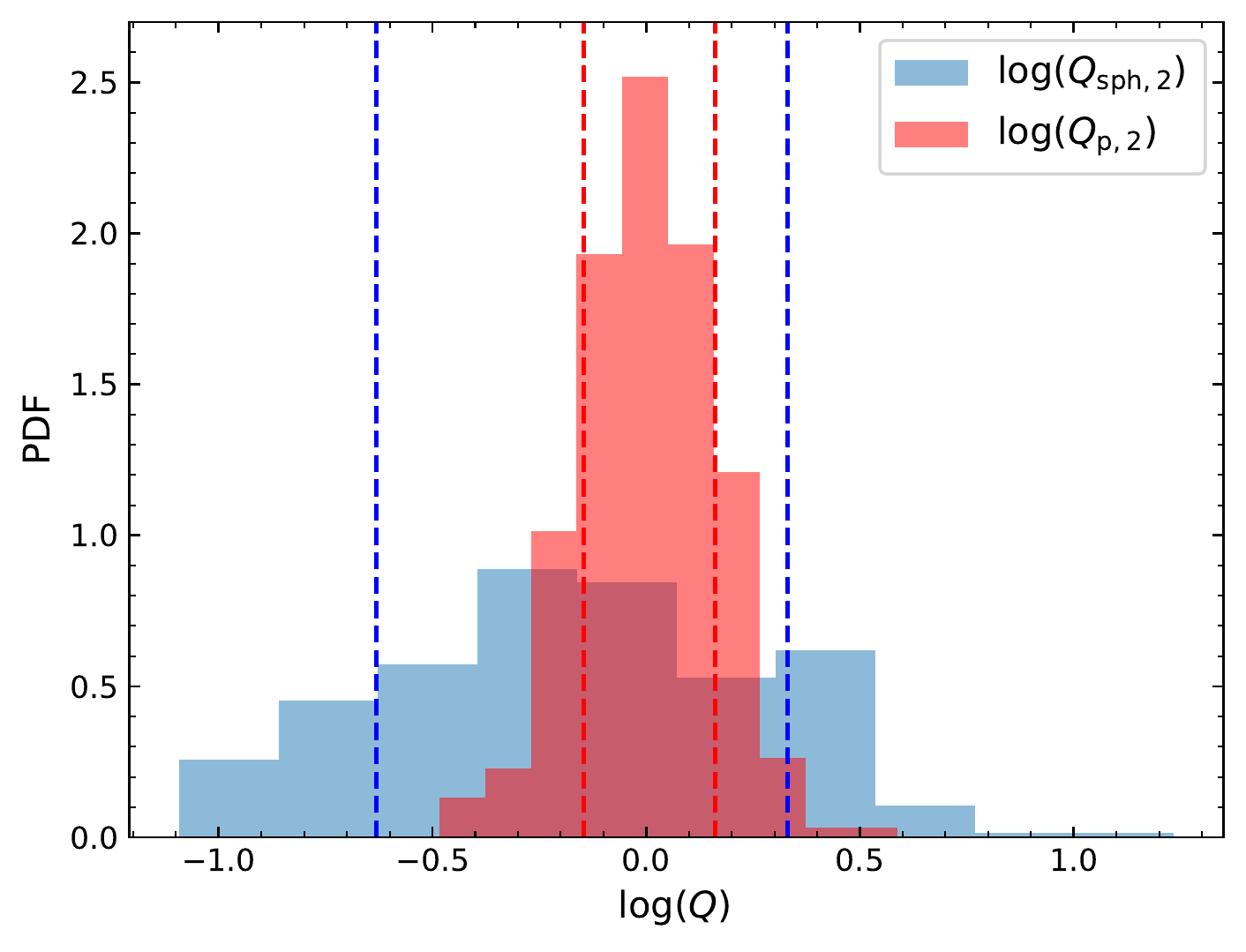}
\caption{Same as \autoref{fig:histogram overlapped}, but now showing the results for effective densities predicted using Fit~2, which includes magnetic field information. Note that the comparison set of $Q_{\rm sph}$ values shown here (blue histogram) is slightly different than that in \autoref{fig:histogram overlapped}, because in this figure we omit the purely hydrodynamic simulations, whereas in the previous figure we included all simulations.}  
\label{fig:histogram overlapped 2}
\end{figure}

To check the effect of including magnetic field data on our density predictions, we perform another MLF on the seven MHD simulations in our data set (hydro and hydroDrive excluded), to which we refer as Fit~2. This fit includes log $B_{\rm pos}$, log $B_{\rm los}$, and $\theta$ (the angle between the plane of sky magnetic field direction and the contour major axis) in the vector of independent variables $\mathbf{X}$. We report the results of this fit in \autoref{tab:MLF results}. We define $\log Q_{\rm p, 2}$ as the logarithm of the ratio of predicted $\rho_{\rm p, 2}$ and effective densities, in analogy with $\log Q_{\rm p}$, and we plot the normalized histograms of $\log Q_{\rm p, 2}$ and $\log Q_{\rm sph}$ in \autoref{fig:histogram overlapped 2}; note that the distribution of $\log Q_{\rm sph}$ shown here is slightly different than that shown in \autoref{fig:histogram overlapped}, since the former includes the contours from hydro and hydroDrive, while this figure excludes them. The dispersion of $Q_{\rm sph}$ for this sample is $\sigma_{\rm sph, 2} = 0.48$~dex, and the dispersion of $Q_{\rm p,2}$ is $\sigma_{\rm p, 2} = 0.15$~dex. Thus, $\Delta\sigma_{2} = \sigma_{\rm sph, 2} - \sigma_{\rm p, 2} = 0.33$~dex. Comparing the results from this and the previous fit, we find fairly minor differences in the fit coefficients and intercepts. The $R^2$ value only increases by about 0.04 from Fit~1 to Fit~2, and $\Delta\sigma_{2}$ is nearly the same as $\Delta\sigma$. This indicates that a model including magnetic field information does not significantly reduce the uncertainty on $\rho_{\rm eff}$ in comparison to one omitting it. Moreover, as summarized in the review by \cite{Crutcher_2012}, magnetic field measurements are observationally expensive: determination of $B_{\rm pos}$ requires measurements of polarised dust continuum emission or absorption, while $B_{\rm los}$ requires Zeeman effect measurements. Due to the long observation times required, these are difficult to obtain for a large sample. Considering the small gains that we have found from including magnetic field information and the difficulty in obtaining it, we generally suggest using Fit~1 to predict the effective density, unless there is magnetic field information available, in which case Fit~2 can be used.

\subsection{Dependence on physical conditions: turbulence, magnetic fields, and outflows}
\label{sec: Examining Fit 1 model under different conditions}

\begin{figure}
   \includegraphics[width = \linewidth]{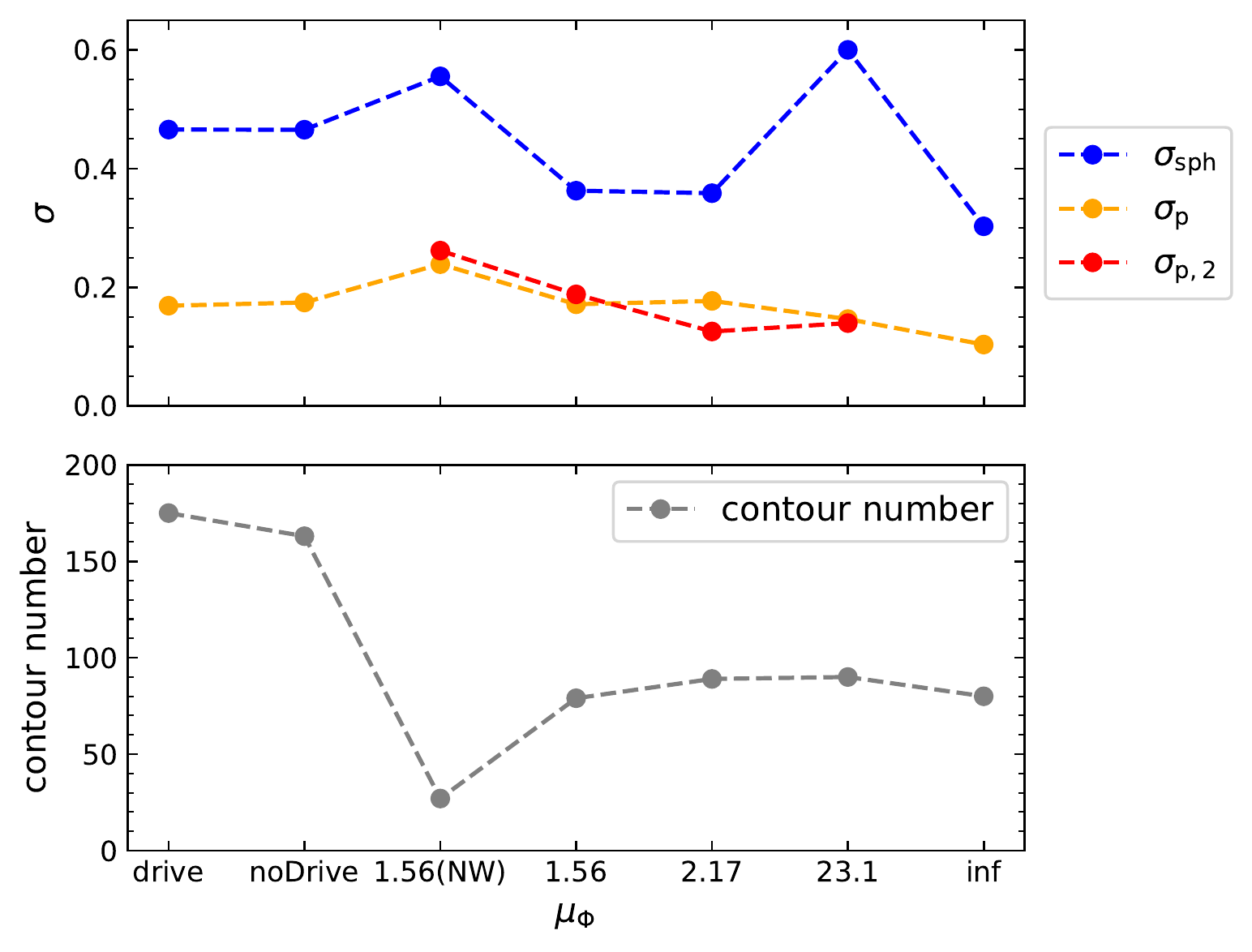}
\caption{Top panel: $\sigma_{\rm sph}$, $\sigma_{\rm p}$ and $\sigma_{\rm p, 2}$ values determined from different subsets of the C18 simulations. Bottom panel: the number of contours selected from different subsets of the C18 simulations. The horizontal axis labels indicate the set of simulations for which these values are measured. Simulation loNW, which has a normalised mass-to-flux ratio $\mu_\Phi = 1.56$ but has protostellar outflows disabled, is only included by itself in set `1.56(NW)'. The `drive' simulation subset includes all four simulations for which turbulent driving continues after gravity is turned on, while `noDrive' includes the other four simulations where there is no driving and turbulence is allowed to decay freely. Each of the last four horizontal axis labels, indicated by numerical values, includes the two simulations (one with and one without driving) with the specified mass-to-flux ratio $\mu_{\Phi}$; here `inf' means $\mu_{\Phi}=\infty$, i.e., the purely hydrodynamic simulations.}  
\label{fig:sim groups}
\end{figure}

We obtain the relation in \autoref{sec: Building the prediction model} by performing MLF on all nine C18 simulations. However, the ambient conditions (mean magnetic field strength, presence or absence of turbulence driving) vary between individual simulations. If the coefficients of the model fit depend on ambient conditions, this may reduce the reliability of our model under specific circumstances. To check whether this is a concern, we use the two linear models developed in \autoref{sec: Building the prediction model} and \autoref{sec: Prediction model with B-field information} to determine the values of $\sigma_{\rm p}$ and $\sigma_{\rm p ,2}$, the dispersions in $\log Q_{\rm p}$ and $\log Q_{\rm p ,2}$, for different subsets of the simulations. We divide the simulations into those with driven versus decaying turbulence, into simulations with different mass-to-flux ratios, and into simulations that do or do not include protostellar outflows. We plot the results in the top panel of \autoref{fig:sim groups}; for comparison we also show $\sigma_{\rm sph}$, the dispersion in $Q_{\rm sph}$ for the same set of simulations. Note that the model obtained via Fit~2 is only applicable for simulation sets excluding hydro and hydroDrive. We provide a full tabulation of the results in \autoref{tab:dispersion results}.

From this plot we can see that for subsets including magnetic fields, there is no significant difference in $\sigma_{\rm p}$ for different models. After applying both prediction models, the dispersion of $Q_{\rm p}$ values is decreased for each subset of the simulations to $\sigma_{\rm p} \in [0.10, 0.26]$~dex; the improvement compared to the simple spherical assumption is in the range $\Delta\sigma \in [0.18, 0.45]$~dex. Therefore, we find relatively little variation in the performance of our prediction model in different simulation subsets; $\sigma_{\rm sph}$ and $\sigma_{\rm p}$ values vary between different sets of simulations, but relatively modestly, so that the errors in the predicted models lie in the range $\approx 0.10-0.26$~dex for each subset of the simulations. Including B-field information brings no improvement from Fit~1 to Fit~2, and the difference is negligible for the weak B-field subset ($\mu_{\Phi} = 23.1$). We do not find a significant correlation between $\sigma_{\rm p}$ and the number of contours (plotted in the bottom panel of \autoref{fig:sim groups}) available for a particular simulation subset. More theoretical work is needed to understand how variations in the ambient conditions affect the relationship between the sky-projected and volumetric quantities, and how they might affect our model. Nonetheless, we can state at this point that the relationship between $\rho_{\rm eff}$ and $\rho_{\rm sph}$ does not seem to depend strongly on the physical conditions present in the star-forming region.

\begingroup
\setlength{\tabcolsep}{10pt} 
\small
\begin{table*}
    \centering
    \begin{tabular}{crrrrrrrr}
    \hline
    Quantity & All & drive & noDrive & 1.56(NW) & 1.56 & 2.17 & 23.1 & inf\\
    \hline
    $\log Q_{\rm sph, 16}$ & $-0.76$ & $-0.59$ & $-0.88$ & $-0.76$ & $-0.33$ & $-0.66$ & $-0.81$ & $-0.83$\\
    $\log Q_{\rm sph, 50}$ & $-0.26$ & $-0.16$ & $-0.48$ & $-0.033$ & $-0.024$ & $-0.34$ & $-0.12$ & $-0.64$\\
    $\log Q_{\rm sph, 84}$ & 0.26 & 0.34 & 0.052 & 0.35 & 0.40 & 0.059 & 0.39 & -0.22\\
    $\sigma_{\rm sph}$ & 0.51 & 0.47 & 0.47 & 0.56 & 0.36 & 0.36 & 0.60 & 0.30\\ \hline
    $\log Q_{\rm p, 16}$ & $-0.16$ & $-0.15$ & $-0.17$ & $-0.30$ & $-0.13$ & $ -0.17$ & $-0.080$ & $-0.19$\\
    $\log Q_{\rm p, 50}$ & $-5.4 \times 10^{-4}$ & $-0.023$ & $0.021$ & 0.071 & 0.049 & $-0.048$ & 0.084 & $-0.093$\\
    $\log Q_{\rm p, 84}$ & 0.18 & 0.19 & 0.18 & 0.18 & 0.21 & 0.18 & 0.21 & 0.017 \\
    $\sigma_{\rm p}$ & 0.17 & 0.17 & 0.17 & 0.24 & 0.17 & 0.18 & 0.15 & 0.10\\ \hline
    $\log Q_{\rm p, 2, 16}$ & --- & --- & --- & $-0.34$ & $-0.16$ & $-0.13$ & $-0.14$ & ---\\
    $\log Q_{\rm p, 2, 50}$ & --- & --- & --- & $-0.068$ & 0.054 & -0.026 & 0.017 & ---\\
    $\log Q_{\rm p, 2, 84}$ & --- & --- & --- & 0.18 & 0.22 & 0.12 & 0.14 & ---\\
    $\sigma_{\rm p, 2}$ & --- & --- & --- & 0.26 & 0.19 & 0.13 & 0.14 & ---\\
    \hline
    \end{tabular}
    \caption{Values of $\log Q_{\rm sph}$, $\log Q_{\rm p}$ and $\log Q_{\rm p,2}$ for different sets of simulations. The first row lists the name of different simulation subsets, where `All' means all simulations and the remaining seven columns correspond to the same subsets of the simulations used in \autoref{fig:sim groups}. In the 1st column, $Q$ is the ratio between the estimated density and the true effective density $\rho_{\rm eff}$, and $\sigma$ is the dispersion of $\log Q$ (Eq~\ref{eq:sigma_sph_definition}). The subscripts 'sph', 'p' and 'p,2' in the 1st column indicate the value of $\log Q$ obtained using the spherical assumption, and the predictive models from Fit~1 and Fit~2, respectively. The subscripts '16', '50', '84' indicate the 16th, 50th and 84th percentile values. Note that Fit~2 is not applicable to simulation sets including hydro or hydroDrive, because those simulations did not include magnetic fields.}
    \label{tab:dispersion results}
\end{table*}
\endgroup

\section{Discussion}
\label{sec: Discussion}

Although our model has proven effective in reducing the uncertainty in observational inferences of $\rho_{\rm eff}$, the physical mechanisms leading to this model are still unclear. In this section, we begin to investigate this question by examining the predictive power of each individual parameter in \autoref{sec: Predicting power of individual parameters}. We then extend our model to account for finite resolution effects in \autoref{sec: Beam size effect}. Finally, we discuss the implications of our findings for observational efforts to measure $\epsilon_{\rm ff}$ and its variation in \autoref{sec: Implications for star formation theories}.

\subsection{Predictive power of individual parameters}
\label{sec: Predicting power of individual parameters}

An obvious question that follows from the success of our MLF model in reducing uncertainties in $\rho_{\rm eff}$ is, which parameters have the most predictive power? We have already seen that magnetic field information adds little accuracy, and we now seek to extend this analysis to the remaining parameters. To investigate this issue, we carry out simple linear fits on the whole sample using only one independent variable each time, and measure the $R^2$ and $\Delta\sigma$ (\autoref{eq:delta_sigma}) values for the fit; the latter characterises the amount by which a model including only that parameter is able to improve estimates of $\rho_{\rm eff}$ relative to the naive spherical assumption. We tabulate the results in \autoref{tab:individual MLF results}. The larger $\Delta\sigma$ is, the more the corresponding parameter can reduce the uncertainty in the effective volume density. The table reveals that the parameters vary widely in their importance. The Gini coefficient $g$ is the most important factor in our model, and by itself it accounts for most of the improvement: $\Delta\sigma = 0.29$~dex for $g$ alone, versus $\Delta\sigma = 0.34$~dex for Fit~1, using all the variables. Next, $\log R_{\rm eff}$, $\log (\sigma_{\rm v, los})$ and $\log (\Sigma_{\rm contour}/\bar{\Sigma})$ have medium predictive power, while the other two parameters have limited influence on the fitted relation.

\begingroup
\setlength{\tabcolsep}{10pt} 
\begin{table*}
    \centering
    \begin{tabular}{ccccc}
    \hline
    Quantity & Intercept & Coefficient & $R^2$ & $\Delta\sigma$ (dex)\\
    \hline
    $g$ & $-0.93$ & 4.6 & 0.75 & 0.29\\
    $\log (R_{\rm eff}/\text{pc})$ & 1.2 & 0.60 & 0.17 & 0.060\\
    $\log (\sigma_{\rm v, los}/\text{(cm/s)})$ & 3.5 & $-0.70$ & 0.099 & 0.051\\
    $\log (\Sigma_{\rm contour}/\bar{\Sigma})$ & $0.0010$ & 0.26 & 0.058 & 0.032\\
    $e$ & $0.25$ & 0.013 & $3.2 \times 10^{-5}$ & 0.0025\\
    $\log (M_*/M_{\rm contour})$ & 0.26 & $-0.0062$ & $6.0 \times 10^{-4}$ & 0.0017\\
    \hline
    \end{tabular}
    \caption{Results of MLF performed on the whole sample with only one independent variable each time. The variables are ranked from top to bottom according to their $\Delta\sigma$ values. For comparison, $\Delta\sigma=0.34$~dex for Fit~1, which uses all six non-magnetic variables.}
    \label{tab:individual MLF results}
\end{table*}
\endgroup

To explain this difference, we need to reexamine \autoref{eq:predicted effective volume density}. Our model is to multiply $\rho_{\rm sph}$ by a correction factor $C$. Thus, if one parameter can reveal how far the object is away from a spherical, uniform-density cloud, then we would expect it to have strong predictive power, or large $\Delta\sigma$. To start with, $g$ describes how concentrated the mass distribution is on the 2D projected map, which is strongly related to the volume-density profile. A larger $g$ corresponds to a larger $\int \rho^{3/2} \, dV$ term and hence a larger $\rho_{\rm eff}$, which is consistent with the positive coefficient of $g$. At the same time, contours with larger $\Sigma_{\rm contour}/\bar{\Sigma}$ and larger $R_{\rm eff}$ might on average be more collapsed along the line of sight, which would suggest a reason for their predictive power: they can flag deviations from the simple spherical assumption. However, the low $R^2$ values of these two individual parameter fits indicate that this is not a strict relation. A contour with larger line-of-sight depth may have larger $v_{\rm los}$ dispersion because of the regions alone the line-of-sight become more uncorrelated, which can explain the medium predictive power of $\sigma_{v}$. However, the lack of correlation between density and velocity dispersion $\sigma_{v}$ has also been found in several observations \citep[e.g.][]{Goodman_2009, Pineda_2008}. \citet{Passot_1998}, \citet{Federrath_2010} and \citet{Federrath_2015} explain this phenomenon as a result of the fact that there is no correlation between density and velocity fluctuations in the case of (near-)isothermal turbulence; though our simulations include stellar radiation feedback, this effect is important only close to protostars, and thus most of the gas is close to isothermal. Therefore, the $R^2$ value of $\log (\sigma_{\rm v, los})$ is also small.

Both other two parameters have limited predictive power.  Similarly, ellipticity may describe how close the 2D contour shape is to a circle, but this apparently provides little constraint on the 3D shape of the gas. Finally, $\log( M_{*}/M_{\rm contour})$ has the smallest $\Delta\sigma$ and $R^2$ values. The reason may be that, once sink particles form in the C18 simulations, the local density profile evolves very little; it likely remains close to the usual $\rho\propto r^{-3/2}$ form expected for free-fall collapse. As a result, the fraction of the available mass that has already accreted, as parameterised by $M_*/M_{\rm contour}$, has very limited predictive power.

Since $g$ is the dominant factor here, we provide a simplified model to predict $\rho_{\rm eff}$ using it alone:
\begin{equation}
    \rho_{\rm p} = 10^{k_g g + b_g} \rho_{\rm sph} = 10^{4.6g - 0.93}\rho_{\rm sph},
    \label{eq: only g model}
\end{equation}
where $k_g$ is the slope and $b_g$ is the intercept from the linear regression. This simplified model can reduce the uncertainty in $\rho_{\rm eff}$ by $\Delta\sigma_{g} = 0.29$~dex. As a consistency check, we note that a spherical cloud with uniform density has surface density Gini coefficient $g_{\rm sph} = 0.2$. Inserting this value into \autoref{eq: only g model} yields $\rho_{\rm p} = 10^{-0.01}\rho_{\rm sph}$, so we would correctly recover $\rho_{\rm p} \approx \rho_{\rm sph}$.

\subsection{Finite resolution effects}
\label{sec: Beam size effect}
\begin{figure}
  \includegraphics[width = \linewidth]{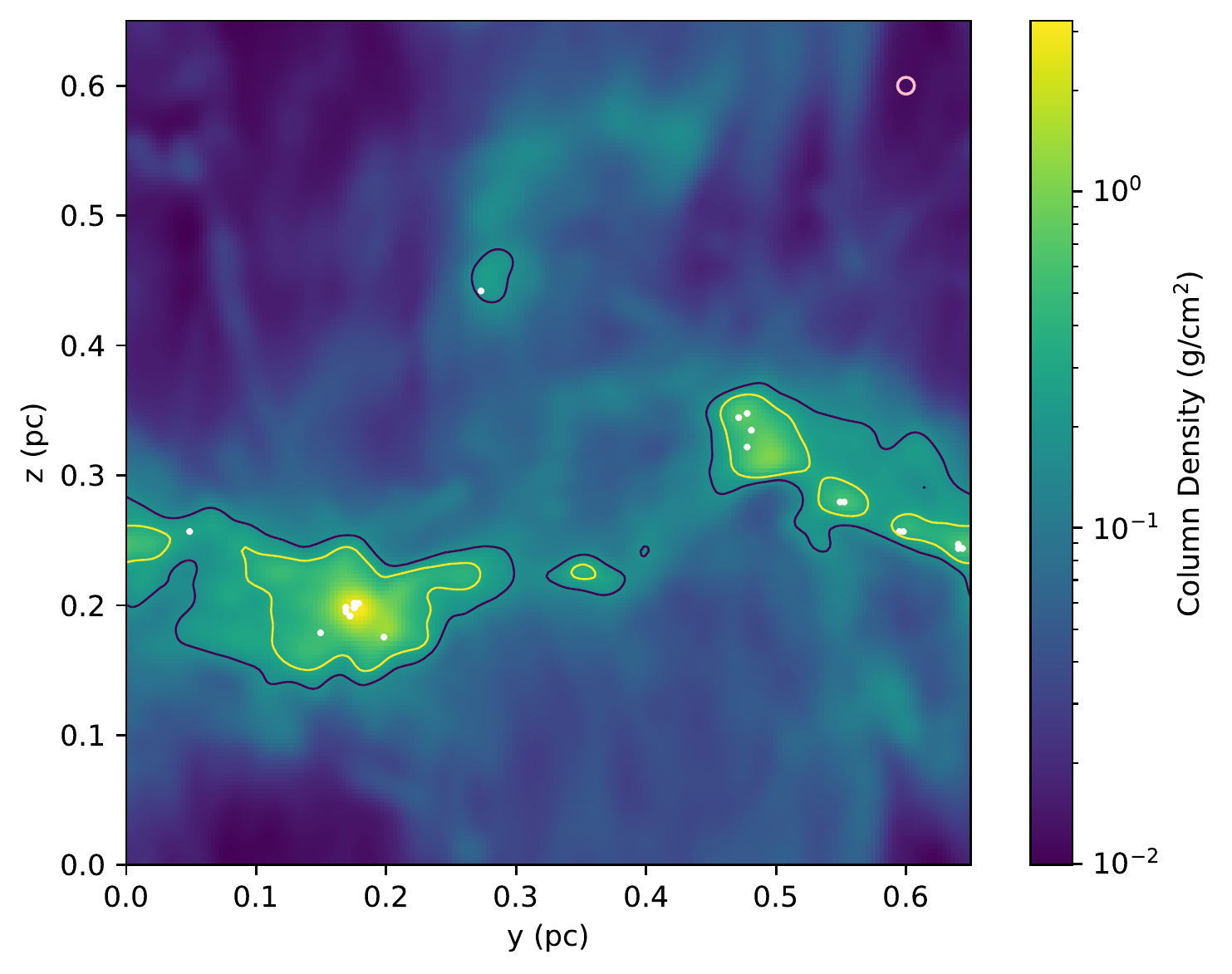}
\caption{The Gaussian-filtered and rebinned column density map of simulation lo, projected along the $\hat{x}$-axis. The size of the Gaussian kernel applied on this map is shown as the pink circle in the right-upper corner. Its radius is $w = L/100$ (note that this is the Gaussian sigma, not the FWHM). The white circles are the projected positions of sink particles. The contours shown represent level~6 ($\Sigma = 0.17 \: \rm g/cm^2 $) and level~11 ($\Sigma = 0.32 \: \rm g/cm^2 $) of the 30~column density levels determined from the map.} 
\label{fig:gaussian map}
\end{figure}

Both the numerical model in \autoref{sec: Building the prediction model} and the simplified model in \autoref{sec: Predicting power of individual parameters} are derived from projection maps created at the native resolution of the simulations, so we are effectively considering only cases where the internal structures of the selected contours are very well-resolved. In real observations the resolution may be limited, and may vary between observations depending on the instrument and the distance to the target. This might have non-trivial effects: a larger beam size will smear details of the contours, and the inferred value of $g$, for example, is very likely to decrease when high-$\Sigma$ peaks are smeared out by low resolution. To explore this effect, we apply a series of Gaussian filters to our projection maps; we consider kernels with standard deviation (\textit{not} full width at half maximum, FWHM) $w$ = $L/1000$, $L/500$, $3L/1000$, $L/250$, $L/200$, $3L/500$, $7L/1000$, $L/125$, $9L/1000$, and $L/100$, where $L$ is the size of the simulation box. We do not consider larger beam sizes because this leaves too small a dynamic range between the size of contours we can resolve and the size scale at which the periodic nature of our simulation box begins to create problems. Then we rebin the Gaussian-filtered maps to a resolution of $2L/w$ pixels on a side, so that the resulting maps are Nyquist-sampled. For each of the rebinned maps, we repeat the analysis presented in \autoref{sec: Data analysis methods}. Note that the 30~contour levels are separately calculated for each rebinned map, and thus are not the same for maps with different levels of beam-smearing, since the contour levels depend on the maximum surface density $\Sigma_{\rm max}$. Similar to \autoref{fig:lo map}, we show a Gaussian-filtered, $\hat{x}$-axis projected column density map of simulation lo in \autoref{fig:gaussian map}. The Gaussian kernel applied on this map is $w= L/100$, which is shown as the pink circle in the right-upper corner. The contours shown are also from level~6 ($\Sigma = 0.17 \: \rm g/cm^2 $) and level~11 ($\Sigma = 0.32 \: \rm g/cm^2 $).

\begin{figure}
  \includegraphics[width = \linewidth]{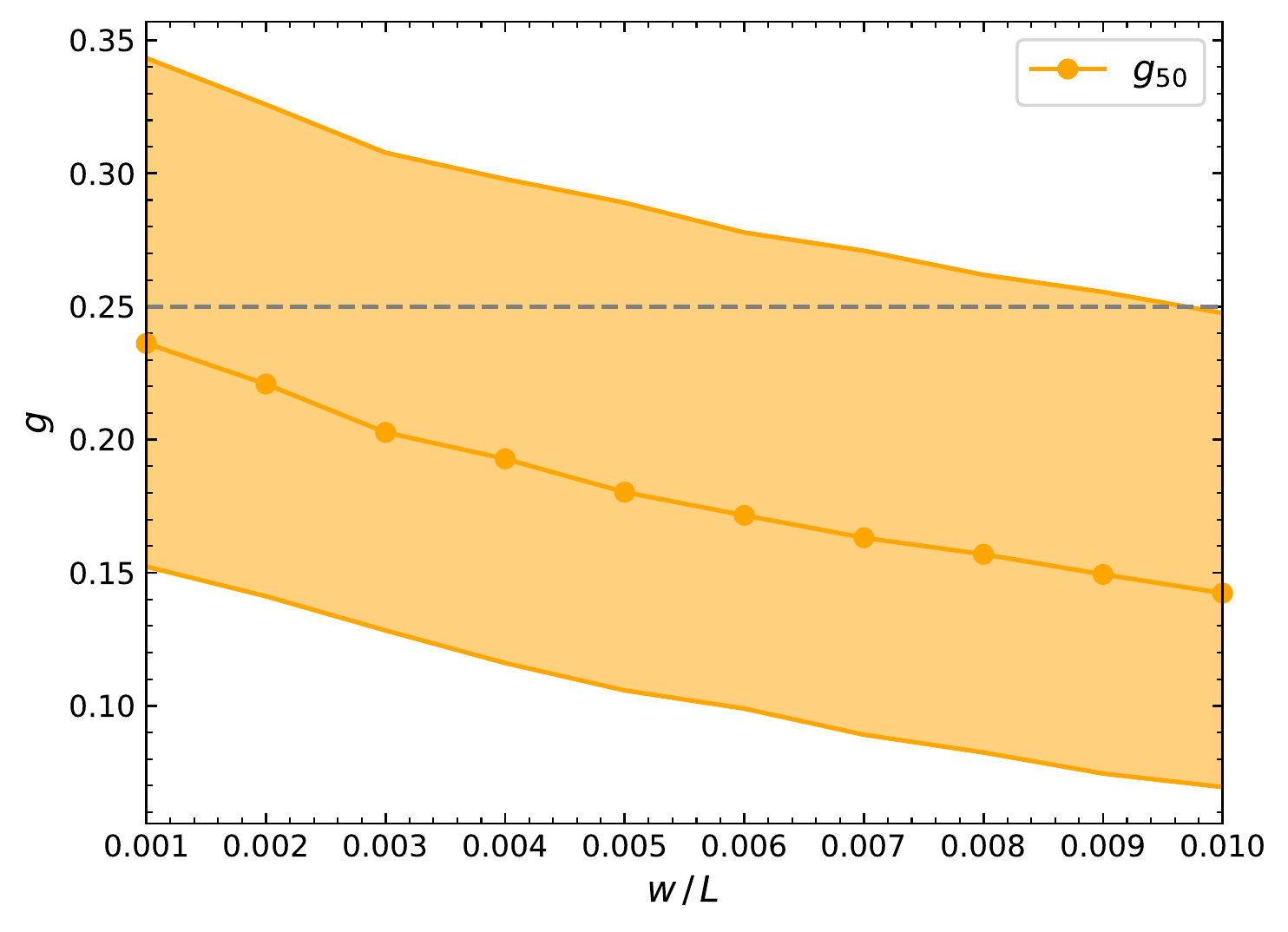}
\caption{The distributions of Gini coefficients computed on the beam-smoothed maps $g$ as a function of smoothing kernel dispersion $w/L$. The upper and lower limits of the band are the 84th and 16th percentiles, while the middle dot points indicate the 50th percentiles. The dashed line is the the median $g$ value $g_{\rm 50, original}$ = 0.25 of the 365 contours selected from original maps.} 
\label{fig:g distribution}
\end{figure}

Since $g$ is the dominant factor in our model and is also likely to be the parameter that is most sensitive to resolution effects, we only study the effect of beam size on the simplified model shown in \autoref{eq: only g model}, which has $g$ as its sole parameter. We begin by investigating the effect of beam size on the values of $g$. We show the distribution of $g$ from selected contours as a function of beam size in \autoref{fig:g distribution}. For $w = L/1000$, we see that the distribution of $g$ is centered around $g = 0.24$, slightly smaller than the median $g$ value $g = 0.25$ of the 365 contours selected from original maps. Larger $w/L$ ratios lead to smaller $g$ values, hence farther from the original distribution. Therefore, the values of $k_g$ and $b_g$ in \autoref{eq: only g model} need to be corrected for the beam size.

\begin{figure}
  \includegraphics[width = \linewidth]{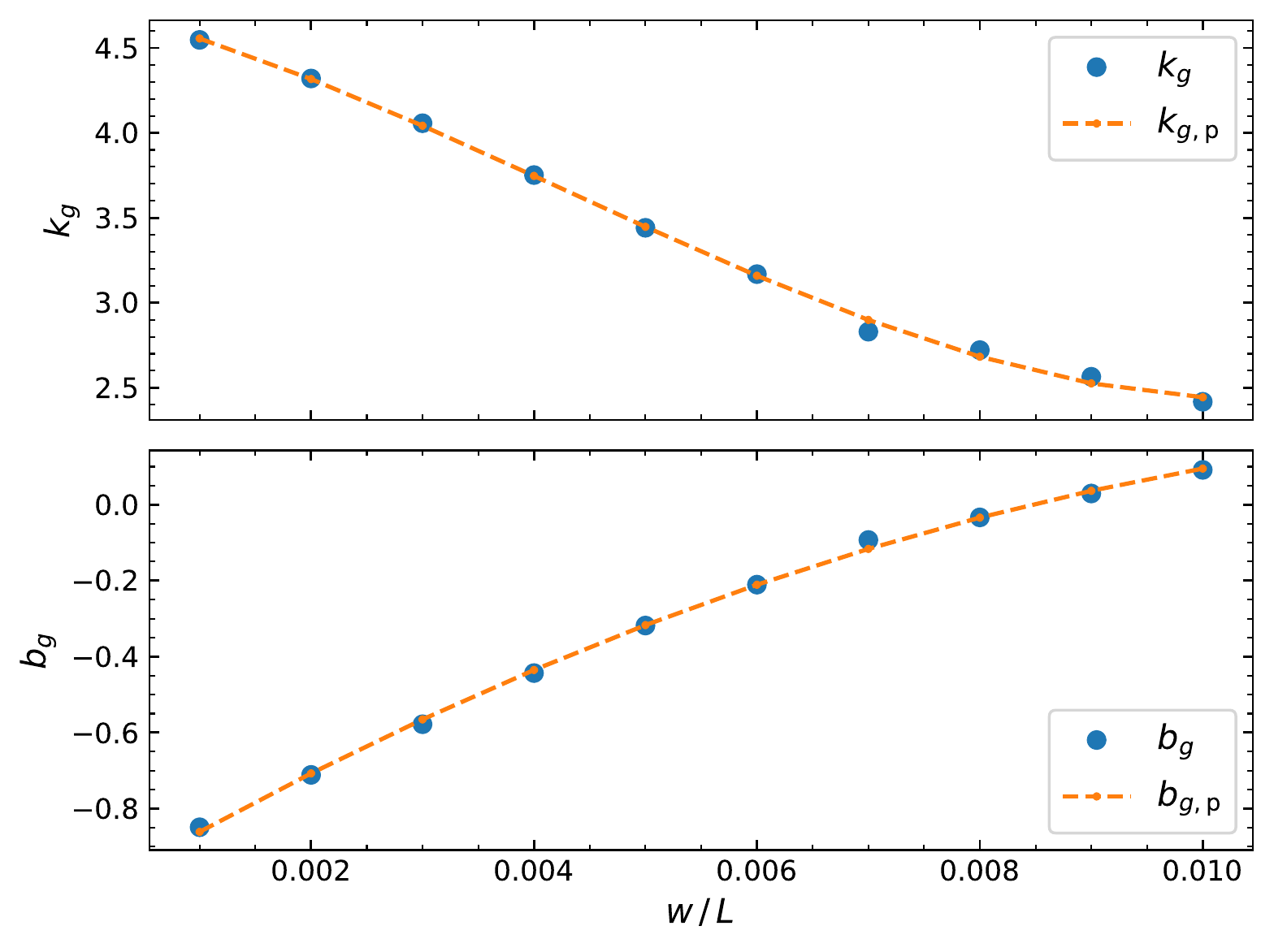}
\caption{Top panel: best-fit coefficient $k_g$ as a function of beam size $w/L$. Bottom panel: same as top panel, but for the intercept $b_g$. In both panels, blue points indicate the numerical results, and orange lines indicate the polynomial fits given by \autoref{eq:k_gp} and \autoref{eq:b_gp}, respectively.} 
\label{fig:k_g and b_g}
\end{figure}

To study how $k_g$ and $b_g$ change with $w/L$, we collect contour properties from maps with the same beam size and then perform linear regressions with only $g$ for each value of $w/L$. We show our best fits for $k_g$ and $b_g$ as a function of beam size in the top and bottom panels of \autoref{fig:k_g and b_g}, respectively. We also show polynomial fits (3rd order for $k_g$, 2nd order for $b_g$) to the results, which capture the variation with high accuracy:
\begin{equation}
    k_{g, \rm p} = 2.7 \times 10^6 \left(\frac{w}{L}\right)^3 - 3.4 \times 10^4 \left( \frac{w}{L}\right)^2 - 1.5 \times 10^2 \left( \frac{w}{L}\right) + 4.7,
    \label{eq:k_gp}
\end{equation}
\begin{equation}
    b_{g, \rm p} = - 6.0 \times 10^3 \left( \frac{w}{L}\right)^2 + 1.7 \times 10^2 \left( \frac{w}{L}\right) - 1.0.
    \label{eq:b_gp}
\end{equation}

\begin{figure}
  \includegraphics[width = \linewidth]{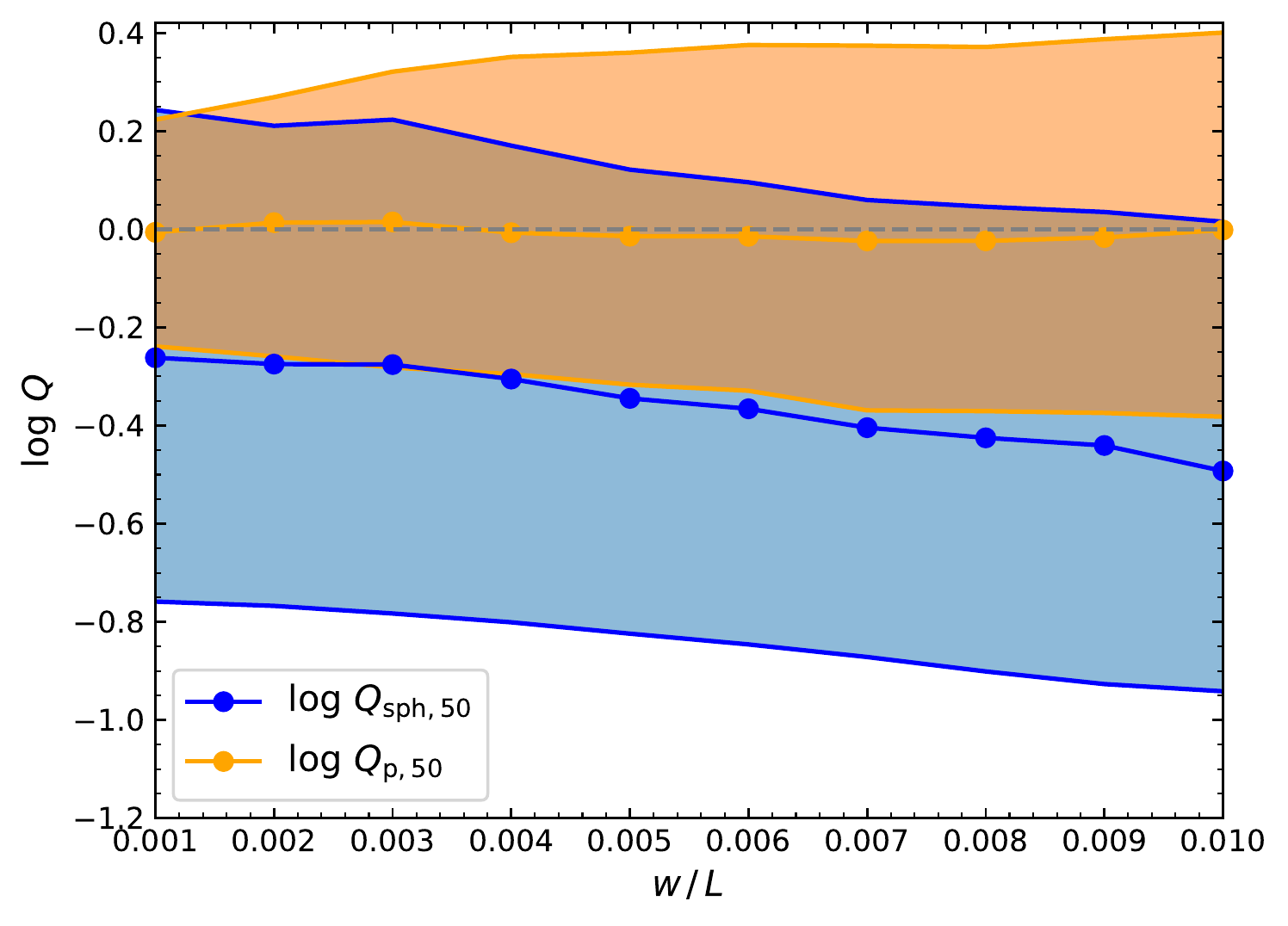}
\caption{The distributions of $\log Q_{\rm sph}$ (blue band) and $\log Q_{\rm p}$ (orange band) versus $w/L$. The upper and lower limits of each band are the 84th and 16th percentiles, while the middle dot plots are the 50th percentiles. The dashed horizontal line indicates $Q=1$, i.e., perfect recovery of $\rho_{\rm eff}$.} 
\label{fig:Qp and Qsph with beam size}
\end{figure}

These fits allow us to predict the effective volume density accounting for beam size effects:
\begin{equation}
    \rho_{\rm p} = 10^{k_{g, \rm p} g + b_{g, \rm p}} \rho_{\rm sph},
    \label{eq: beam and g model}
\end{equation}
where $k_{g, \rm p}$ and $b_{g, \rm p}$ are determined by \autoref{eq:k_gp} and \autoref{eq:b_gp}. The distributions of $\log Q_{\rm sph}$ and $\log Q_{\rm p}$ resulting from this procedure are shown in \autoref{fig:Qp and Qsph with beam size}. This plot reveals several interesting conclusions. First, $\log Q_{\rm sph, 50}$ is centred around $-0.26$ for highly-resolved observations ($w/L = 0.001$, i.e., $\sim 1000$ resolution elements across the molecular cloud), and drops for lower resolution. This means that $\rho_{\rm sph}$ calculated in observations will underestimate $\rho_{\rm eff}$, which leads to an overestimate of $\epsilon_{\rm ff}$. This bias will be increased for poorly-resolved observations. The offset in $\rho_{\rm eff}$ can be as large as $-0.49$~dex when $w/L = 0.01$, corresponding to a systematic overestimate of $\epsilon_{\rm ff}$ by $\approx 0.25$~dex. Our predictive model corrects this systematic error, so $\log Q_{\rm p}$ is centred around 0, independent of beam size, with a maximum offset of only 0.015~dex. The predictive model also continues to reduce the dispersion in $\rho_{\rm eff}$ estimates, though the improvement $\Delta\sigma$ decreases from 0.27~dex at high resolution to 0.087~dex at the coarsest resolution we consider. This degradation in performance is not surprising, since we have access to less information about the internal density structure of objects in the coarser observations. In summary, our correction model, \autoref{eq: beam and g model}, can both eliminate the resolution-dependent offset between $\rho_{\rm sph}$ and $\rho_{\rm eff}$ and reduce the uncertainty of $\rho_{\rm sph}$, which can can greatly enhance the accuracy of $\epsilon_{\rm ff}$ measurements. 

\subsection{Implications for previous measurements of $\epsilon_{\rm ff}$, and for star formation theories}
\label{sec: Implications for star formation theories}
As shown in \autoref{fig:Qp and Qsph with beam size},$\rho_{\rm sph}$ underestimates $\rho_{\rm eff}$, which leads to a systematic overestimate of $\epsilon_{\rm ff}$; a simple linear fit to our results gives
\begin{equation}
    \Delta\epsilon_{\rm ff} = -0.5 \; \text{log} \; Q_{\rm sph, 50} = 13\frac{w}{L} + 0.11\; \text{dex},
    \label{eq: epsilon_ff vs beam}
\end{equation}
where $w$ is the resolution and $L$ is the map size. To examine the possible $\Delta\epsilon_{\rm ff}$ caused by beam size effects in observations, we take the example of the $\epsilon_{\rm ff}$ study by \cite{Ochsendorf2017}. They use the Magellanic Mopra Assessment (MAGMA) DR3 \citep{Wong_2011} CO intensity map to determine molecular could mass in the Large Magellanic Cloud (LMC), which has a beam size of 45" FWHM and a map size of 3.6 deg$^2$. Inserting these factors into \autoref{eq: epsilon_ff vs beam} predicts $\Delta\epsilon_{\rm ff} = 0.16$ dex, which is a relatively small offset, and smaller than the scatter determined by \cite{Ochsendorf2017} as $\sigma_{\epsilon_{\rm ff}} \approx 0.4$~dex. This result suggests that the possible overestimation of $\epsilon_{\rm ff}$ may not be significant in observations. This result, however, needs further investigation since \autoref{eq: epsilon_ff vs beam} is fitted with the fixed simulation domain size $L$, which is not the exact equivalent of the observed map size in a real galaxy. We discuss this issue further in \autoref{sec: future work}. Nonetheless, this result suggests that the bias in $\epsilon_{\rm ff}$ measurements due to finite resolution is not a severe effect.

However, it is not only the mean value of $\epsilon_{\rm ff}$ that is crucial for theories of star formation. Its spread, $\sigma_{\epsilon_{\rm ff}}$, is also important, because theoretical models predict widely differing values of $\sigma_{\epsilon_{\rm ff}}$. For example, \cite{Lee_2016} calculate $\sigma_{\epsilon_{\rm ff}}$ values for different theoretical models, predicting values of 0.24~dex for the turbulence-regulated model of \cite{2005ApJ...630..250K} and 0.12 or 0.13~dex for the multi-free-fall model of \cite{Hennebelle2011}, depending on the choice of parameters. Models in which $\epsilon_{\rm ff}$ increases with time as a cloud evolves give larger dispersions: $\sigma_{\epsilon_{\rm ff}} = 0.54$~dex for $\epsilon_{\rm ff}\propto t$ \citep{Murray15, Lee15}, and 0.9~dex for $\epsilon_{\rm ff}\propto t^2$ \citep{Feldmann_2010}. In observations of Milky Way molecular clouds that use the spherical approximation to determine $\epsilon_{\rm ff}$ \citep[e.g.,][]{2013ApJ...778..133L, Evans_2014, Heyer2016}, $\sigma_{\epsilon_{\rm ff}}$ is estimated to be about 0.35~dex, which is significantly larger than the spread predicted by the first two models, and much smaller than the value expected from the time-dependent models.

\autoref{sec: Comparing rho_eff and rho_sph} suggests a somewhat different interpretation, however: there we show that $\rho_{\rm sph}$ typically differs from $\rho_{\rm eff}$ by $\sigma_{\rm sph} \approx  0.51$~dex, so even if $\epsilon_{\rm ff}$ were perfectly constant in reality, a measurement of it that relies on the spherical assumption would be expected to show a dispersion $\sigma_{\epsilon_{\rm ff},\rm sph} \sim 0.26$~dex. Conversely, the intrinsic scatter in $\epsilon_{\rm ff}$ suggested by an observed dispersion of 0.35~dex is $\sigma_{\epsilon_{\rm ff}, \rm intrinsic} \approx \sqrt{0.35^2 - 0.26^2}$ = 0.23~dex. This result directly casts doubt on the star formation models predicting larger $\epsilon_{\rm ff}$ scatters. It suggests that a significant part of the observed scatter is not reflective of true scatter in $\epsilon_{\rm ff}$, but instead represents observational error induced by reliance on the spherical assumption. This conclusion is consistent with the analysis of \citet{Krumholz_2020}, who argue based on statistical modelling of star clusters and pre-cluster gas clumps that the intrinsic spread in $\epsilon_{\rm ff}$ must be substantially smaller than the observed spread.

\section{Sample application to the Ophiuchus Cloud}
\label{sec: Analysis on Ophiuchus Cloud}

To test the effectiveness of our simplified Gini model, \autoref{eq: only g model}, on real data, we study the SFEs of regions in the Ophiuchus cloud. The observations we use are described by \citet{Pokhrel_2020}, and we refer readers to that paper for full details of data processing. To summarise the most important points here: \citet{Pokhrel_2020} obtain a map of the H$_2$ column density $N(\rm H_2)$ from the Herschel Gould Belt Survey (HGBS) archive \citep{Andr__2010}, and they combine this with a catalogue of young stellar objects (YSOs) drawn from the Spitzer Extended Solar Neighborhood Archive (SESNA) compiled by R. Gutermuth et al. (in preparation). The Ophiuchus cloud $N(\rm H_2)$ map has a pixel size of $d_{\rm oph}$ = 0.002 pc, which can be converted into a Gaussian filter standard deviation $w_{\rm oph} = d_{\rm oph}/1.18 = 0.0017$ pc. As the cloud size is 11.5 $\times$ 12.0 pc$^2$,  the $w/L$ ratio is $\approx 10^{-4}$. Finite resolution effects are therefore very limited, and we can just apply \autoref{eq: only g model}.

The first step in our analysis is to create and select contours on different column density levels. Following \citet{Pokhrel2021}, we define 106 $N(\rm H_2)$ levels linearly spaced between $2.82 \times 10^{21}$ cm$^{-2}$ and $5.22 \times 10^{22}$ cm$^{-2}$. We then choose contours for further analysis according to our three selection conditions. First, we discard contours with no YSO inside. Second, we choose contours with mean radius no less than 30 pixels ($\approx$ 0.06 pc) to guarantee their internal structures are well resolved. Third, for the remaining contours on each level, we only select the most massive one. After selection, we have 75 contours as the observation sample.

\begin{figure}
  \includegraphics[width = \linewidth]{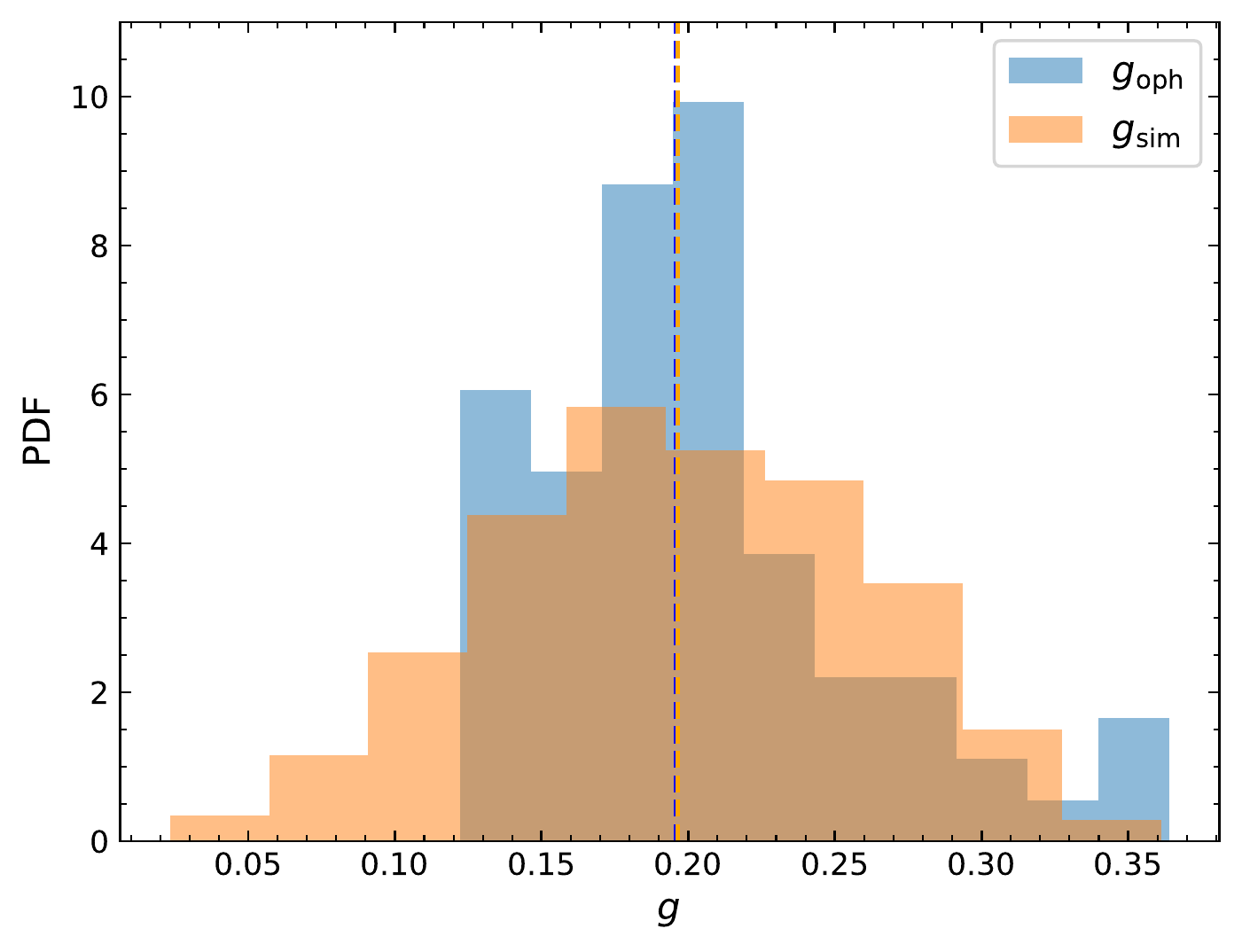}
\caption{Histograms of the distribution of Gini coefficients from the Ophiuchus cloud, $g_{\rm oph}$ (blue), and from the C18 simulations, $g_{\rm sim}$ (orange). Both distributions have been normalised to have unit integral. The dashed vertical lines show the median values of the two distributions.} 
\label{fig:g_oph and g_sim distributions}
\end{figure}

As an initial check of our method, we wish to verify that the distributions of $g$ from the simulations and observation are similar. This comparison requires some care. \citet{Pokhrel_2020} mask pixels for which their analysis returns an estimated column density $N({\rm H}_2) > 10^{23}$ cm$^{-2}$, because at these high column densities the cloud may be optically thick in one or more of the \textit{Herschel} bands; consequently, the values they derive represent only lower limits. The range between the observed mean column density $\bar{N}(\rm H_2) = 3.40 \times 10^{21}$ cm$^{-2}$ and the highest unmasked value is only 1.47 dex, while this range in the x-projection map from simulation hi, for example, is 2.71 dex. In order to make a fair comparison between simulations and observations, we must clip the simulations so their dynamic range is comparable to that of the observations. Thus for each projection map from C18 simulations we mask pixels with $\Sigma > 10^{1.47} \bar{\Sigma}$, and repeat the contour selection process described in \autoref{sec: Creating and selecting contours}. We then determine $g$ for these new contours from the C18 simulations, $g_{\rm sim}$, and compare to the distribution of Gini coefficients in the observed map, $g_{\rm oph}$, in \autoref{fig:g_oph and g_sim distributions}. The two distributions are clearly qualitatively similar, and the median values of the two samples are nearly identical: $g_{\rm oph, med} = 0.196$ and $g_{\rm sim, med} = 0.197$. A two-sided Kolmogorov-Smirnov test comparing the two samples returns a $p$ value of $p = 0.18$, indicating that we cannot rule out the null hypothesis that these two $g$ samples were drawn from the same parent distribution. Therefore, we conclude that the $g$ distributions from the Ophiuchus cloud and C18 simulations are consistent with one another.

\begin{figure}
  \includegraphics[width = \linewidth]{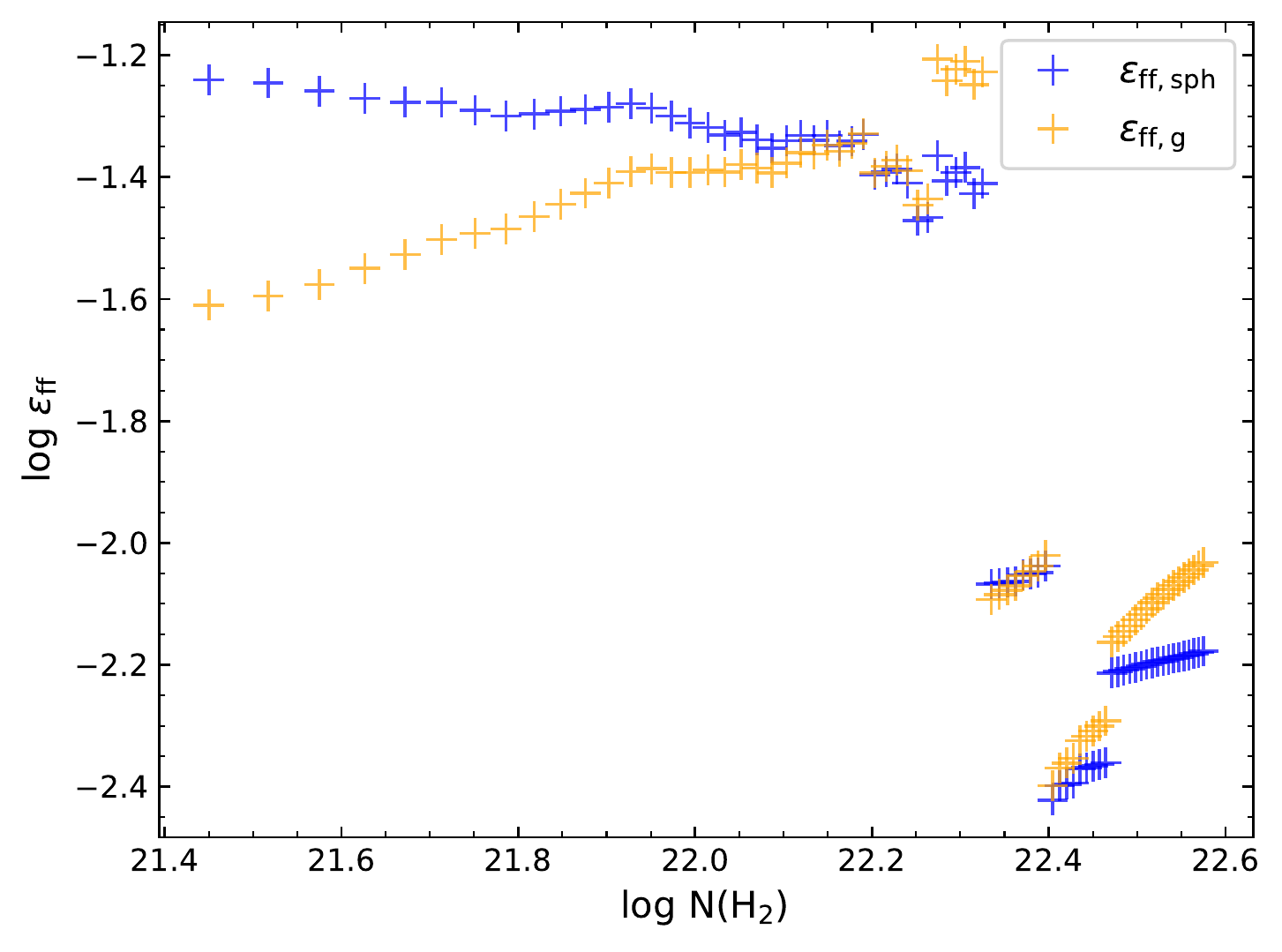}
\caption{Star formation efficiencies of the selected contours from Ophiuchus cloud. The x-axis is the $\log N({\rm H}_2)$ level at which the contour is selected, and $\epsilon_{\rm ff, oph}$ (blue) and $\epsilon_{\rm ff, g}$ (orange) are the star formation efficiencies determined using the spherical assumption (\autoref{eq:rhosph}) and using our Gini model (\autoref{eq: only g model}), respectively.
} 
\label{fig: SFE oph plot}
\end{figure}

\begin{figure}
  \includegraphics[width = \linewidth]{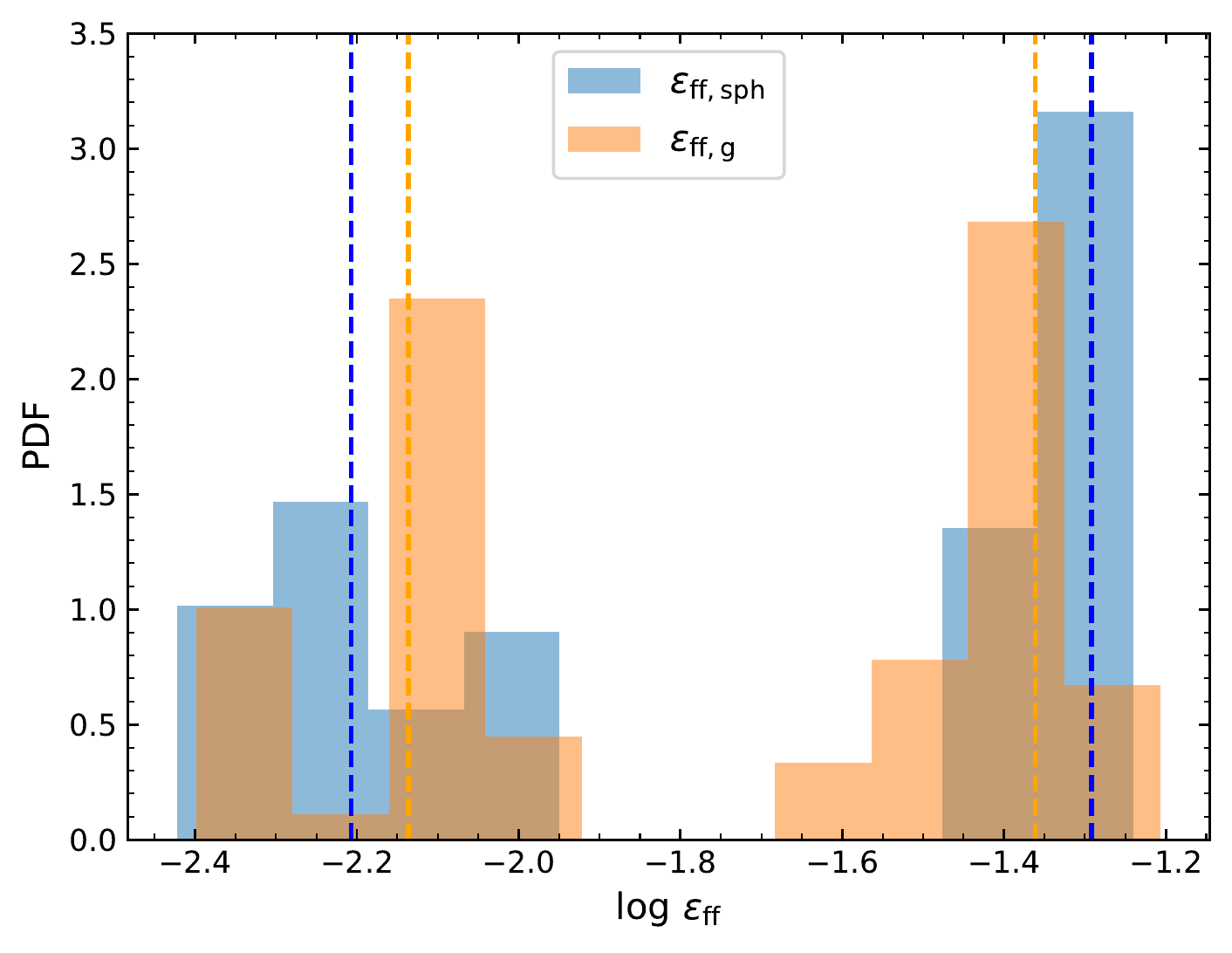}
\caption{Normalized histograms of $\epsilon_{\rm ff, sph}$ (blue) and $\epsilon_{\rm ff, g}$ (orange). The blue and orange dashed vertical lines are the 16th and 84th percentiles of the two distributions, respectively.
} 
\label{fig: SFE oph hist}
\end{figure}

We next determine the SFEs of the Ophiuchus cloud contours. For every contour, we measure the enclosed gas mass $M_{\rm gas}$,  the enclosed area $A$, and the number of enclosed protostars $N_{\rm PS}$. We compute the SFR $\dot M_*$ of one contour as
\begin{equation}
   \dot M_* = N_{\rm PS}M_{\rm PS}/t_{\rm PS},
    \label{eq:SFR}
\end{equation}
where $M_{\rm PS} \approx 0.5 M_{\rm \odot}$ is the mean mass of protostars in our catalogue \citep{Evans_2009}, and $t_{\rm PS} \approx 0.5 Myr$ is the duration of the protostellar phase during which YSOs will be included in this catalogue \citep{Dunham_2015}. We determine the mean volume density in two ways: one using \autoref{eq:rhosph} (the spherical assumption) and one using \autoref{eq: only g model} (our Gini model). With these values we can determine the SFEs with \autoref{eq:star formation efficiency}. We plot the resulting values of $\epsilon_{\rm ff}$ as a function of contour level $N({\rm H}_2)$ in \autoref{fig: SFE oph plot}. The sudden drop in $\epsilon_{\rm ff}$ at the high column density is probably due to the YSOs moving out of the contours during the protostar stage \citep{Pokhrel2021}. Comparing the results of the two methods of estimating the density, we find that applying our Gini model has the effect of shifting the high and low ends of the $\epsilon_{\rm ff}$ distribution towards the middle. We show this more clearly in \autoref{fig: SFE oph hist}, which shows the distributions of $\epsilon_{\rm ff}$ derived with the two density estimation methods, together with their 16th and 84th percentiles. The median values we obtain with the spherical and Gini methods of density estimation are $\log \epsilon_{\rm ff, sph, med} = -1.4$ and $\log \epsilon_{\rm ff, g, med} = -1.5$, respectively, and the dispersions of $\epsilon_{\rm ff, sph}$ and $\epsilon_{\rm ff, g}$ are $\sigma_{\rm sph} = 0.46$~dex and $\sigma_{\rm g} = 0.39$~dex, respectively. Thus using the Gini method to estimate the volume density decreases the estimated dispersion of SFE inside the Ophiuchus cloud by $\Delta\sigma = 0.07$~dex. This is smaller than the $0.5 \Delta\sigma_g = 0.15$~dex found in our idealised tests. However, our idealised tests did not include the effects of limited dynamic range (which are likely qualitatively similar to the effects of beam smearing); moreover, this result is from contours inside one single cloud, while a conclusion can only be drawn by studying several molecular clouds. Nevertheless, the fact that we find $\Delta\sigma > 0$ is an encouraging result for our model.

\section{Future work}
\label{sec: future work}

Although our predictive model has proven its ability to reduce the uncertainty of effective volume density estimates, there is still much room for future improvement. The first step would be to enlarge the sample with data from different simulations. Although the C18 simulations capture many of the physical processes and conditions in dense, star-forming molecular clouds, and span a very wide range of physical parameters (magnetic field strength, turbulent driving), they still have several limitations. For example, they apply purely solenoidal turbulent driving, whereas  in reality both solenoidal modes from galactic differential rotation and compressive modes from stellar feedback may be present \citep{Federrath_2018IAU, Federrath_2018}. Another limitation is from their radiative transfer methods. They assume the gas and the dust share the same temperature. This assumption of strong coupling is valid at densities above $\sim 10^4 - 10^5$ $\rm cm^{-3}$ \citep{Goldsmith_2001}, but may fail for lower density, non-self-gravitating regions, which leads the simulations to overestimate the dust cooling rate for the gas. If we were to extend our analysis to other simulations without these limitations, we might extend the range of our contour sample and obtain better fits. 

Another potential area of improvement is the fitting method. Our current MLF method is justified by its high $R^2$ results, but the resulting model is highly dominated by $g$. Moreover, the variables used in the MLF may not be completely independent of each other. A contour with small $R_{\rm eff}$, for example, is more likely to have large $\bar{\Sigma}$ because we are focusing on the centre of a molecular cloud. A linear relation, in this case, may not be the ideal form, and we should explore the possibility of other forms of correlations. If we were able to enlarge the sample size with more simulations, one possible approach would be to utilize machine learning to discover the underlying relations.

In \autoref{sec: Beam size effect} we use the ratio between the beam size and map size $w/L$ for analysing the effects of beam-smearing. Expressing the results in terms of $w/L$ has the advantage that it makes the results dimensionless. However, the simulated cloud size is actually infinite because of the periodic boundary condition applied in the C18 simulations, while $L$ is only the simulation domain size and should neither be seen as the equivalent of a molecular cloud size nor as a projection map size in observations. Real molecular clouds have edges, and our simulations do not. Since this problem originates from the simulations themselves, we probably cannot overcome it using the C18 data. Instead, a better approach would be to start from galactic-scale simulations, form molecular clouds self-consistent within them, and continue zooming in until we reach the dense clump scale often used in $\epsilon_{\rm ff}$ estimates. This would provide a sample of simulated molecular clouds with well-defined physical sizes, from which we could derive relations for beam-size effects more comparable to observations.

We have tested our Gini model on the observation data of Ophiuchus cloud. To obtain more conclusive results of SFE and $\sigma_{\rm SFE}$, however, one need to study several different molecular clouds. Meanwhile, besides the resolution effect, the effect of protostars shifting out of contours and the large error of column density in dense regions should also be considered. Our current plan is to conduct a survey on the 12 molecular clouds studied in \cite{Pokhrel_2020}, whose results may put more regulations on theoretical star formation models. 

\section{Conclusion}
\label{sec: Conclusion}

This work aims at obtaining precise measurements of the star formation efficiency of molecular clouds. Making these measurements requires that we estimate the volume densities of gas clouds seen only in projection; these estimates are a major source of error, and reducing them is the primary goal of this work. We use a suite of simulations of star formation from \cite{Cunningham2018} to investigate the nature of this error. We first evaluate the effect of assuming that the clouds we see are spherical and uniform density, the most common approach in the current literature. Then we develop a numerical model that can predict the effective volume density of a projected 2D contour from its observable properties substantially more accurately than the simple spherical assumption. We build this model with multiple linear fitting, and the high coefficient of determination we obtain ($R^2\sim 0.83$) demonstrates that this produces reliable results.

We find that the volume density determined from the spherical assumption has a significant scatter $\sigma_{\rm sph} = 0.51$~dex, and a underestimation $log Q_{\rm sph, med} = 0.26$~dex, compared to the true, free-fall time weighted mean density, which is the quantity of interest for measurements of the star formation efficiency. Considering these effects, the star formation efficiencies determined in recent studies relying on the spherical assumption are likely to be overestimated by 0.13~dex, and the scatter $\sigma_{\epsilon_{\rm ff}}\sim 0.35$~dex, likely represents a true, intrinsic scatter in the star formation efficiency of no more than 0.23~dex, imposing strong constraints on theoretical models.

By comparison, when we apply our linear model, using all the observable parameters we tested, we reduce the uncertainty of the mean density by as much as $\Delta\sigma = 0.34$~dex. We also evaluate the influence of individual parameters in our predictive model, and suggest physical explanations of their significance and relative predictive power. In cases where we observe only the mass, area, column density, and the Gini coefficient of a target cloud, a simplified model can still decrease the uncertainty by $\Delta\sigma = 0.29$~dex. This improvement is sufficient to roughly halve the uncertainties of recent star formation efficiency measurements, and thus is very substantial. The effectiveness of this simplified model is proven by our analysis of the Ophiuchus cloud. In addition, we investigate the effect of the telescope beam size on our simplified model and provide a corrected version to minimize this effect.

Despite its good performance, this model still has much room for future development. We can extend its applicable range by including more simulations spanning a larger variety of physical conditions. Rebuilding the model with machine learning may also enhance its capabilities.

\section*{Acknowledgements}
We would like to thank Prof. Andrew J. Cunningham for sharing the C18 simulation data. M.~R.~K.~acknowledges funding from the Australian Research Council (Discovery Project DP190101258 and Future Fellowship FT180100375), and the Australia-Germany Joint Research Cooperation Scheme (UA-DAAD). C.~F.~acknowledges funding provided by the Australian Research Council (Discovery Project DP170100603 and Future Fellowship FT180100495), and the Australia-Germany Joint Research Cooperation Scheme (UA-DAAD). R.~P. and R.~A.~G. acknowledge funding support for this work from NASA ADAP awards NNX15AF05G, 80NSSC18K1564 and NNX17AF24G. R.~P. acknowledges funding support from NASA ADAP award 80NSSC18K1564, and R.~A.~G. acknowledges funding support from NASA ADAP awards NNX11AD14G and NNX13AF08G.  We further acknowledge high-performance computing resources provided by the Australian National Computational Infrastructure (grants~jh2 and ek9) in the framework of the National Computational Merit Allocation Scheme and the ANU Merit Allocation Scheme, and by the Leibniz Rechenzentrum and the Gauss Centre for Supercomputing (grant~pr32lo).

This research has made use of data from the Herschel Gould Belt survey (HGBS) project \footnote{\url{http://gouldbelt-herschel.cea.fr}}. The HGBS is a Herschel Key Programme jointly carried out by SPIRE Specialist Astronomy Group 3 (SAG 3), scientists of several institutes in the PACS Consortium (CEA Saclay, INAF-IFSI Rome and INAF-Arcetri, KU Leuven, MPIA Heidelberg), and scientists of the Herschel Science Center (HSC).

\section*{Data Availability}

The simulation and observation data underlying this article will be shared upon reasonable request to the corresponding author.



\bibliographystyle{mnras}
\bibliography{MyLibrary} 


\bsp	
\label{lastpage}
\end{document}